\pdfoutput=1
\documentclass[journal]{IEEEtran}
\usepackage{amsfonts}
\usepackage{bm}
\usepackage{amsmath,amsfonts,amsthm,amssymb}
\usepackage{setspace}
\usepackage{Tabbing}
\usepackage{fancyhdr}
\usepackage{lastpage}
\usepackage{extramarks}
\usepackage{chngpage}
\usepackage{algorithmic}
\usepackage{soul,color}
\usepackage{epsfig}
\usepackage{graphicx,float,wrapfig,subfigure}
\usepackage{dsfont}
\usepackage{longtable}
\usepackage{wrapfig}
\usepackage{stfloats}
\usepackage{cite}
\usepackage{graphicx}
\usepackage{array}
\usepackage{multirow}{\tiny }
\usepackage{multicol}
\usepackage{colortbl}
\usepackage{tabularx}
\usepackage{mdwmath}
\usepackage{mdwtab}
\usepackage{color}
\usepackage{verbatim}
\usepackage{amsmath}
\usepackage{tikz}
\usetikzlibrary{calc}
\usepackage{flowchart}
\usetikzlibrary{shapes.geometric, arrows}
\usepackage{overpic}
\usepackage{epstopdf}
\usepackage{url}
\usepackage{bbm}
\usepackage[colorlinks,linkcolor=black,anchorcolor=black,citecolor=black,urlcolor=black]{hyperref} 

\usepackage{makecell}
\usepackage{textcomp}
\usepackage{algorithm}
\usepackage{algorithmic}

\usepackage{threeparttable}

\newtheorem{proposition}{Proposition}

\newtheorem{remark}{Remark}

\usepackage{color}
\makeatletter %
\let\myorg@bibitem\bibitem
\def\bibitem#1#2\par{%
	\@ifundefined{bibitem@#1}{%
		\myorg@bibitem{#1}#2\par
	}{%
		\begingroup
		\color{\csname bibitem@#1\endcsname}%
		\myorg@bibitem{#1}#2\par
		\endgroup
	}%
}

\makeatother %

\allowdisplaybreaks

\usepackage{footnote}
\makesavenoteenv{tabular}
\makesavenoteenv{table}

\begin{document}

\title{
Scheduling HVAC loads to promote renewable generation integration with a learning-based joint chance-constrained approach
	}

\author{
Ge~Chen,~\IEEEmembership{Graduate Student Member,~IEEE,}
Hongcai~Zhang,~\IEEEmembership{Member,~IEEE,}
Hongxun~Hui,~\IEEEmembership{Member,~IEEE,}
and~Yonghua~Song,~\IEEEmembership{Fellow,~IEEE}
\vspace{-8mm}

}

\maketitle

\begin{abstract}
The integration of distributed renewable generation (DRG) in distribution networks can be effectively promoted by scheduling flexible resources such as heating, ventilation, and air conditioning (HVAC) loads. However, finding the optimal scheduling for them is nontrivial because DRG outputs are highly uncertain. To address this issue, this paper proposes a learning-based joint chance-constrained approach to coordinate HVAC loads with DRG. Unlike cutting-edge works adopting individual chance constraints to manage uncertainties, this paper controls the violation probability of all critical constraints with joint chance constraints (JCCs). This joint manner can explicitly guarantee the operational security of the entire system based on operators' preferences. To overcome the intractability of JCCs, we first prove that JCCs can be safely approximated by robust constraints with proper uncertainty sets. A famous machine learning algorithm, one-class support vector clustering, is then introduced to construct a small enough polyhedron uncertainty set for these robust constraints. A linear robust counterpart is further developed based on the strong duality to ensure computational efficiency. Numerical results based on various distributed uncertainties confirm the advantages of the proposed method in optimality and feasibility.

\end{abstract}
\begin{IEEEkeywords}
Demand-side flexibility, joint chance constraints, support vector clustering, HVAC systems, renewable energies
\end{IEEEkeywords}

\section{Introduction} \label{sec_intro}
\IEEEPARstart{T}{he} increasing penetration of renewable generation mitigates the demand for fossil energy. However, it also exacerbates the imbalance between demands and generation due to the stochastic characteristics of renewable generation \cite{8661899}.
As a result, much distributed renewable generation (DRG) has to be curtailed to maintain operational security. Fortunately, the integration of DRG can be promoted by properly scheduling flexible sources in distribution networks \cite{8779816}.
Due to the natural ability of buildings to store heating/cooling power, heating, ventilation, and air conditioning (HVAC) loads can be regarded as desirable demand-side flexible sources \cite{8606271,9151228}. Hence, tons of scheduling methods have been proposed to realize the proper coordination between HVAC loads and DRG. For instance, reference \cite{8592000} proposed an aggregation method to identify the operational flexibility of multiple buildings. Reference \cite{9535415} developed a simplification method to model the flexibility of large-scale HVAC systems. Reference \cite{9072559} regarded HVAC loads as virtual battery systems and provided a control scheme to unlock their flexibility. 

However, it is hard to predict DRG outputs due to their stochastic characteristics perfectly. Thus, uncertainties from forecasting errors of DRG must be considered during the power scheduling to avoid constraint violations \cite{8859599}. Robust optimization has traditionally been used to manage these uncertainties \cite{9862574}. However, it is usually overly conservative because every constraint should be satisfied with all possible realizations of uncertainties. Chance constrained programming (CCP) is an alternative method to manage the uncertainties from DG in OPF \cite{geng2019data}. It allows constraint violations with a small probability so that operators can effectively balance robustness and optimality based on their preferences. Reference \cite{8591956} proposed a CCP-based framework to coordinate thermostatically controlled loads with uncertain DRG. Reference \cite{wei2020multi} leveraged CCP to handle the uncertainties from DRG during the planning of integrated energy systems. Reference \cite{8110715} developed a multi-period CCP framework to aggregate HVAC loads for reserve services. Reference \cite{8254387} utilized a distributionally robust CCP method to schedule flexible loads and energy storage systems in distribution networks. Reference \cite{9417102} proposed a fast distributionally robust CCP framework to utilize the load flexibility of HVAC loads for cost minimization. However, most CCP-based methods, including \cite{wei2020multi,8110715,8254387,9417102}, control the violation probability of critical constraints individually. For a distribution system, violations of any constraint may affect operational safety. Therefore, the system security is hard to guarantee for this individual manner because it does not restrict all critical constraints to be satisfied simultaneously with a predetermined probability \cite{7973099}.

The violation probabilities of critical constraints should be jointly considered due to the security concern of power systems \cite{9122389,chen2022deep}. Therefore, joint chance constraints (JCCs), which jointly describe these probabilities, are preferable for coordinating HVAC systems with DRG in distribution networks. Unfortunately, JCCs are much more intractable than individual ones \cite{yu2018copula}. Hence, much effort has been made to overcome this. Generally speaking, the published methods for handling JCCs can be divided into three categories. 

\subsubsection{Bonferroni approximation}
The Bonferroni approximation leverages multiple individual chance constraints (ICCs) to approximate an intractable JCC. By requiring that the summation of the risk parameters in all ICCs is smaller than the original one in the JCC, the flexibility of solutions can be guaranteed \cite{odetayo2018chance}. For instance, references \cite{hassan2018optimal} employed the Bonferroni approximation to handle the JCCs in optimal power flow problems. References \cite{baker2019joint} provided tighter upper bounds for the Bonferroni approximation to improve its optimality. However, the risk parameter in each ICC is very small because of the flexibility requirement. Thus, the corresponding solution is usually overly conservative, especially when the number of jointly considered constraints is large \cite{bertsimas2018data}. 

\subsubsection{Scenario approach}
The scenario approach approximates JCCs by requiring constraints to be robust for a finite number of scenarios. With a sufficient number of scenarios, the feasibility of the original JCC can be guaranteed \cite{geng2019data}. Reference \cite{8017474} utilized the scenario approach to approximate JCCs with scenario-wise deterministic scenarios. References \cite{8060613, 8626040} designed box uncertainty sets to cover all scenarios and proposed a robust optimization-based inner approximation to reduce computational burdens. Other types of uncertainty sets, e.g., ellipsoids \cite{8357484} or convex hulls \cite{8706676}, have also been combined with the scenario approach to improve its optimality. However, the approach may still be overly conservative once some extreme samples of uncertainties are chosen as scenarios.

\subsubsection{Sample average approximation}
The sample average approximation (SAA) utilizes enough empirical samples to approximate the underlying true distribution of uncertainties. SAA introduces auxiliary binary variables to identify safe samples (satisfying all constraints) and unsafe samples (causing constraint violations). By restricting the total number of unsafe samples, JCCs can be approximated by sample-wise deterministic constraints. Note that SAA differs from the scenario approach by accepting small constraint violations. Hence SAA can potentially achieve better optimality \cite{geng2019data}. SAA has been utilized to handle JCCs in power grid planning \cite{8680681} and unit commitment \cite{7822944}. However, since SAA needs to introduce many binary variables, it is computationally expensive.

This paper aims to design an energy- and time-efficient scheduling method to overcome the challenges above, i.e., overly conservative results or huge computational burdens. To achieve this goal, a novel scheduling approach is proposed. Its main contributions are threefold:
\begin{enumerate}
    \item We propose a joint chance-constrained model to manage the uncertainties from DRG during the coordination of HVAC loads and DRG in distribution networks. The violation probabilities of all critical constraints are jointly considered so that the system security can be explicitly guaranteed with a high probability. 
	\item We design a robust approximation with a novel learning-based uncertainty set to replace intractable JCCs. By leveraging the one-class support vector clustering (OC-SVC) algorithm, we establish a small polyhedron uncertainty set to tightly cover most historical samples for an energy-efficient solution. We further prove that this uncertainty set can guarantee the feasibility of solutions. To the best of our knowledge, this is the first adoption of the OC-SVC-based uncertainty set for joint chance-constrained scheduling problems. 
	\item We develop a tractable robust counterpart for the OC-SVC-based uncertainty set. This counterpart is linear and can be easily solved by off-the-shelf solvers with guaranteed computational efficiency.
\end{enumerate}

The rest of this paper is  organized as follows. Section \ref{sec_formulation} provides the system model. Section \ref{sec_solution} presents the   procedure of the proposed method. Section \ref{sec_case} discusses simulation results, and section \ref{sec_conclusion} presents the conclusion.  

\section{Problem formulation} \label{sec_formulation}
HVAC systems and DRG are usually connected to distribution networks. By properly designing the power schedule of HVAC systems, the integration of uncertain DRG can be promoted to reduce the total cost of distribution networks. Indoor thermal discomfort should be avoided to guarantee the service quality of HVAC systems. The scheduling strategy should satisfy bus voltage and branch power flow limitations to ensure the system's operational security.

\subsection{Modeling of building thermal dynamics}
The indoor temperature variation in buildings can be described by a widely-used energy conservation model,
\begin{align}
\theta_{i,t}^\text{in}=&a_i^\text{in}\theta_{i,t-1}^\text{in}+a_i^\text{out}\theta_{t-1}^\text{out} + a_i^\text{h}h_{i, t-1} \notag \\ &+ a_i^\text{q} p_{i,t-1}^\text{HV}, \  \forall i \in \mathcal{I}, \  \forall t \in \mathcal{T}, \label{eqn_thermal}
\end{align}
where $\theta_{i,t}$ and $\theta_{t-1}^\text{out}$ are the respective temperatures of the indoor and outdoor environments; $h_{i, t}$ is the heat load contributed by indoor sources (e.g., humans and electric devices); $p_{i,t}^\text{HV}$ is the active power of the HVAC system; and
\begin{align}
\begin{cases}
a_i^\text{in}=e^{-\frac{\Delta t}{R_i C_i}}, \quad a_i^\text{out}=1-a_i^\text{in}, \\
a_i^\text{h}=R_i \cdot a_i^\text{out}, \  a_i^\text{q} = - \text{COP}_i \cdot a_i^\text{h},
\end{cases}  \forall i \in \mathcal{I},
\end{align}
where $C_i$ is the heat capacity of the building connected to the $i$-th node, $R_i$ is the corresponding thermal resistance between indoor and outdoor environments, and $\Delta t$ is the length of the optimization time interval; $\text{COP}_i$ is the coefficient of performance of the HVAC system.
The reactive power of the HVAC system calculated by
\begin{align}
q_{i,t}^\text{HV} = \sqrt{(1 - \phi_i^2)}/\phi_i \cdot p_{i,t}^\text{HV}, \ \forall i \in \mathcal{I}, \forall t \in \mathcal{T}, \label{eqn_q_HV}
\end{align} 
where $\phi_i$ is the power factor of the $i$-th HVAC system. Due to thermal comfort requirements and device limitations, the indoor temperature and active power of HVAC systems are bounded by
\begin{align}
\underline{\bm \theta} \leq \bm \theta_{t}^\text{in} \leq \overline{\bm \theta}, \quad \bm p_t^\text{HV} \leq \overline{\bm p}, \quad \forall t \in \mathcal{T}, \label{eqn_comfort}
\end{align}
where $\underline{\bm \theta}$ and $\overline{\bm \theta}$ are the lower and upper bounds, respectively, of the comfortable range; $\overline{\bm p}$ is the upper limitation of the active power of HVAC systems. 

\subsection{Modeling of distribution networks}
\subsubsection{Operation cost}
The energy cost of a distribution network is the cost of purchasing electricity (the net load at the substation is positive) plus the profit from selling power to the upper-level grid (net load is negative),
\begin{align}
&EC_t = (\eta^\text{buy} G_t^\text{buy} - \eta^\text{sell} G_t^\text{sell})\Delta t,\quad \forall t \in \mathcal{T}, \label{eqn_EC}\\
&G_t^\text{buy} - G_t^\text{sell} = G_t^\text{g}, \quad G_t^\text{buy}\geq0,\quad G_t^\text{sell}\geq0,\quad \forall t \in \mathcal{T}. \label{eqn_netload}
\end{align} 
where $EC_t$ is the energy cost at time $t\in\mathcal{T}$; $\eta^\text{buy}$ and $\eta^\text{sell}$ are the per-unit prices of electricity purchasing and selling, respectively, $G_t^\text{buy}$ and $G_t^\text{sell}$ are two auxiliary variables; $G_t^\text{g}$ is the net power at the substation; $\Delta t$ is the length of a time slot. 

\subsubsection{Nodal power injections}
The nodal power injections can be expressed as\footnote{While we assume DRG only outputs active power here, the proposed model can be readily extended to consider a DG's reactive power output.}
\begin{align}
&\bm p_t = -\bm p_{t}^\text{HV} - \bm p_{t}^\text{e} + \bm G_t^\text{DG}*\bm \lambda_t, \  0 \leq \bm \lambda_t \leq 1,\  \forall t \in \mathcal{T}, \label{eqn_p} \\
&\bm q_t = - \bm q_{t}^\text{HV} - \bm q_{t}^\text{e},\quad \forall t \in \mathcal{T}, \label{eqn_q}
\end{align}
where $\bm p_{t}^\text{HV}$ and $\bm q_{t}^\text{HV}$ are vector forms of $p_{i,t}^\text{HV}$ and $q_{i,t}^\text{HV}$, respectively; $\bm G_t^\text{DG}$ and $\bm \lambda_t$ are the available output and utilization rate of DRG, respectively; $\bm p_{t}^\text{e}$ and $\bm q_{t}^\text{e}$ are the active and reactive base loads, respectively (i.e., the loads apart from the power demands of HVAC systems); "$*$" denotes element-wise multiplication. In practice, the available output of DRG is usually uncertain, which can be expressed as
\begin{align}
\bm G_t^\text{DG} = \overline {\bm G}_t^\text{DG} * (1+\bm \xi_t),\quad \forall t \in \mathcal{T}, \label{eqn_uncertain_DG}
\end{align}
where $\overline {\bm G}_t^\text{DG}$ and $\bm \xi_t$ are the nominal available power and uncertain level of DRG, respectively. 

\subsubsection{Power flow constraints}
Buildings with flexible HVAC loads are spatially distributed, so the coordination strategy should satisfy power flow constraints, i.e., bus voltage and branch power flow limitations. The power flow in a distribution network can be described by the linearized DistFlow \cite{wang2014coordinated},
\begin{align}
\begin{cases}
P_{ij,t} = \sum_{k\in \mathcal{C}_{j}} P_{jk,t} - p_{j,t}, \\
Q_{ij,t} = \sum_{k\in \mathcal{C}_{j}} Q_{jk,t} - q_{j,t},\\
U_{j,t}=U_{i,t}-2(r_{ij}P_{ij,t}+x_{ij}Q_{ij,t}), \\
\forall (i,j) \in \mathcal{B},\ \forall t \in \mathcal{T},
\end{cases} \label{eqn_distflow}
\end{align}
where $P_{ij,t}$ and $Q_{ij,t}$ are the active and reactive power flows, respectively, on branch $(i, j)\in\mathcal{B}$;  $i$ and $j$ are the bus indexes, $i,j\in \mathcal{I}$; $p_{j,t}$
and $q_{j,t}$ are the active and reactive power injections, respectively, at bus $j$; $U_i$ is the square of the voltage at bus $i$; $r_{ij}$ and $x_{ij}$ are the resistance and reactance, respectively, of branch $(i,j)$; and set $\mathcal{C}_{j}$ contains the child bus indexes of bus $j$, $\mathcal{C}_{j}\subseteq \mathcal{I}$. The net power at substation $G_t^\text{g}$ is  the summation of the active power flows from the slack bus ($i=0$) to its child buses,
\begin{align}
G_t^\text{g} = \sum_{j\in \mathcal{C}_{0}} P_{0j,t},\quad\forall t \in \mathcal{T}.\label{eqn_netload2}
\end{align} 

All bus voltages and branch power flows should lie in a proper range to ensure system security. According to  (\ref{eqn_p}) and (\ref{eqn_q}), uncertainty $\bm \xi_t$ affects the power injections on each bus. Based on the linearized Distflow (\ref{eqn_distflow}), this uncertainty further propagates to the square of the bus voltage $\bm U_t$, and branch power flows $\bm P_t$ and $\bm Q_t$. To maintain  system security, a JCC is employed to manage the uncertainties:
\begin{align}
\mathbb{P}\left(
\begin{array}{l}
U_{i, \text{min}} \leq {U}_{i, t} \leq U_{i, \text{max}}, \ \forall i \in \mathcal{I}, \\
P_{ij,t}^2 + Q_{ij,t}^2 \leq (S_{ij}^\text{max})^2,  \forall (i,j) \in \mathcal{B}, 
\end{array}\right) \geq 1 - \epsilon, \label{eqn_JCC_ori}
\end{align}
where $U_{i, \text{min}}$ and $U_{i, \text{max}}$ are the lower and upper bounds of the bus voltage; $S_{ij}^\text{max}$ is the maximum allowable apparent power flow. Symbol $\epsilon$ is the risk parameter, which defines the maximum allowable probability of constraint violations. 
\begin{remark}
Here, we use JCCs instead of individual ones to control the violation probability of all security constraints under the impacts of uncertainties from DRG. This is preferable in the coordination of DRG and HVAC loads because it can accurately ensure the security of the entire system with a predetermined confidence level.
\end{remark}

\subsubsection{Linearization of the quadratic JCC}
There are quadratic constraints inside the probability operator of (\ref{eqn_JCC_ori}), which are intractable. To address this issue, we introduce auxiliary variables $P_{ij,t}^\text{aux}$ and $Q_{ij,t}^\text{aux}$, and conservatively approximate each quadratic constraint inside the probability operator by
\begin{align}
&\begin{cases}
(P_{ij,t}^\text{aux})^2+(Q_{ij,t}^\text{aux})^2 \leq (S_{ij}^\text{max})^2, \\
|P_{ij,t}| \leq P_{ij,t}^\text{aux},\ |Q_{ij,t}| \leq Q_{ij,t}^\text{aux}.
\end{cases} \forall (i,j) \in \mathcal{B}. \label{eqn_extract}
\end{align}
Note that $P_{ij,t}^\text{aux}$ and $Q_{ij,t}^\text{aux}$ are deterministic. Thus, the second-order cone constraint, i.e., the first line of (\ref{eqn_extract}), does not contain any uncertainty. Then, by substituting (\ref{eqn_extract}), the quadratic JCC (\ref{eqn_JCC_ori}) can be approximated by a deterministic second-order cone constraint plus a linear JCC
\begin{align}
&(P_{ij,t}^\text{aux})^2+(Q_{ij,t}^\text{aux})^2 \leq (S_{ij}^\text{max})^2, \quad \forall (i,j) \in \mathcal{B}, \label{eqn_SOCP}\\
&\mathbb{P}\left(
\begin{array}{l}
U_{i, \text{min}} \leq {U}_{i, t} \leq U_{i, \text{max}}, \ \forall i \in \mathcal{I},\\
|P_{ij,t}| \leq P_{ij,t}^\text{aux},|Q_{ij,t}| \leq Q_{ij,t}^\text{aux},\forall (i,j) \in \mathcal{B},
\end{array}\right) \notag \\ 
&\quad \quad\quad\quad\quad\quad\quad\quad\quad\quad \quad\quad \geq 1-\epsilon, \ \forall t \in \mathcal{T}. \label{eqn_JCC}
\end{align}
Obviously, any feasible solution of  (\ref{eqn_JCC}) must be feasible for  (\ref{eqn_JCC_ori}). Hence  (\ref{eqn_JCC}) can guarantee the feasibility of solutions.

\subsection{Optimization problem}
Our goal is to minimize the total cost. Thus, the  optimization problem is formulated as
\begin{align} 
&\min_{(\bm p_t^\text{HV}, \bm \lambda_t, \bm P_{t}^\text{aux}, \bm Q_{t}^\text{aux})_{\forall t \in \mathcal{T}}} \quad \mathbb{E} \left(\sum_{t \in \mathcal{T}} EC_t \right) \tag{$\textbf{P1}$},\\
&\begin{array}{r@{\quad}r@{}l@{\quad}l}
\text{s.t.} &\text{Eqs. } &\text{(\ref{eqn_thermal})-(\ref{eqn_netload2}) and   (\ref{eqn_SOCP})-(\ref{eqn_JCC})}.
\end{array} \notag
\end{align}

\section{Learning-based reformulation for joint chance constraints} \label{sec_solution}
In \textbf{P2}, constraint (\ref{eqn_JCC}) is intractable because of the inside probability operator. We address this issue through a learning-based approach. We first propose a robust approximation for the JCC to eliminate the intractable probability operator. Then, an OC-SVC-based method to construct a tight polyhedron uncertainty set for the previous robust approximation. A linear counterpart for the proposed OC-SVC-based uncertainty set is further developed to ensure computational tractability.

\subsection{Robust approximation of the JCC} \label{sec_RobustApproximation}
Observing that all constraints inside the probability operator of (\ref{eqn_JCC}) are linear, they can be expressed in generic form
\begin{align}
(\bm H_{m,t} \bm \xi_{t})^\intercal \bm y_{t} \leq \beta_{m,t}(\bm y_{t}), \forall m \in \mathcal{M}, \forall t \in \mathcal{T},
\end{align}
where $\bm H_{m,t}$ and $\bm \xi_t$ are the coefficient matrix and random variable, respectively, in each constraint; $\bm y_{t}=[\bm p_t^\text{HV}, \bm \lambda_t, \bm P_{t}^\text{aux}, \bm Q_{t}^\text{aux}]$ is the decision variable vector;   $\beta_{m,t}$ is an affine function of $\bm y_{t}$; and $m \in \mathcal{M}=\{1,2,\cdots,M\}$ is the index of the constraints inside the probability operator. By defining a new function $h_t(\bm y_{t}, \bm \xi_{t})$ as
\begin{align}
   h_t(\bm y_{t}, \bm \xi_{t}) = \max_{m \in \mathcal{M}} \big((\bm H_{m,t} \bm \xi_{t})^\intercal \bm y_{t} - \beta_{m,t}(\bm y_{t})\big), \forall t \in \mathcal{T}, \label{eqn_h_RO}
\end{align}
Eq. (\ref{eqn_JCC}) can be expressed as
\begin{align}
    \mathbb{P}\left(h_t(\bm y_{t}, \bm \xi_{t}) \leq 0 \right) \geq 1 - \epsilon, \quad \forall t \in \mathcal{T}. \label{eqn_JCC_generic}
\end{align}
If we can find an uncertainty set $\mathcal{U}_t$   satisfying
\begin{align}
\mathbb{P} \left(\bm \xi_t \in \mathcal{U}_t\right) \geq 1-\epsilon, \forall t \in \mathcal{T}, \label{eqn_U_require}
\end{align}
the intractable JCC (\ref{eqn_JCC_generic}) can be approximately expressed by a robust constraint,
\begin{align}
\max_{\bm \xi_t \in \mathcal{U}_t} \left(h(\bm y_t, \bm \xi_t) \right)\leq 0, \forall t \in \mathcal{T}, \label{eqn_RO}
\end{align}
and the probability operator can be   eliminated. Obviously, the key problem of this robust approximation is to find an appropriate $\mathcal{U}_t$, which should satisfy   feasibility, optimality, and tractability requirements:
\begin{enumerate}
\item \textbf{Feasibility}: $\mathcal{U}_t$ must cover at least $100(1-\epsilon)\%$ of samples in the historical dataset, i.e., Eq. (\ref{eqn_U_require});
\item \textbf{Optimality}: The space covered by $\mathcal{U}_t$ should be small, so that the obtained solution can achieve desirable energy efficiency;
\item \textbf{Tractability}: The robust approximation (\ref{eqn_RO}) with $\mathcal{U}_t$ as its uncertainty set must have a tractable counterpart so that it can be effectively handled by off-the-shelf solvers. 
\end{enumerate}

\subsection{OC-SVC-based uncertainty set} \label{sec_unsupervised}
We leverage a popular learning-based anomaly detection method, one-class support vector clustering (OC-SVC), to construct the appropriate uncertainty set $\mathcal{U}_t$ \cite{shang2017data}. 
For convenience, we omit the subscript $t$   in this section. 
The   idea of OC-SVC is to find a minima sphere in high-dimension feature space as the boundary to separate   normal and abnormal data, which can be expressed as
\begin{align}
&\min_{R, \bm o, \bm \omega} \quad R^2+\frac{1}{N \epsilon}\sum_{n \in \mathcal{N}} \omega_n, \tag{$\textbf{P-SVC}$}\\
&\begin{array}{r@{\quad}r@{}l@{\quad}l}
\text{s.t.} && \Vert \bm \phi(\bm \xi^{(n)}) - \bm o \Vert^2 \leq R^2+\omega_n, \quad \forall n \in \mathcal{N},
\end{array} \label{eqn_SVC_c1}\\
&\quad \quad \quad\omega_n\geq 0, \quad \forall n \in \mathcal{N}, \label{eqn_SVC_c2}
\end{align}
where $R$ is the radius of the sphere; $\epsilon$ is a controllable parameter; $\bm \phi(\cdot)$ is a mapping function from the original space to high-dimensional feature space; $\bm o$ is the center of the sphere; and
$\omega_n$ is a slack variable. The second term in the objective function is the penalty for outliers. By introducing dual variables $\bm \alpha$ and $\bm \beta$ for (\ref{eqn_SVC_c1}) and (\ref{eqn_SVC_c2}), the KKT conditions of $\textbf{P-SVC}$ are
\begin{align}
&\bm 1^\intercal \bm \alpha = 1, \quad \bm o = \sum_{n \in \mathcal{N}}\alpha_n \bm \phi(\bm \xi^{(n)}), \quad \bm \alpha + \bm \beta = \frac{1}{N \epsilon} \cdot \bm 1, \label{eqn_KKT1} \\
&\omega_n \beta_n = 0,\quad \forall n \in \mathcal{N},\label{eqn_KKT2}\\
&\alpha_n\left(R^2+\omega_n-\Vert \bm \phi(\bm \xi^{(n)}) - \bm o \Vert^2\right)=0, \quad \forall n \in \mathcal{N},\label{eqn_KKT3}
\end{align}
where (\ref{eqn_KKT1}) is obtained by setting the derivative of the Lagrangian function to zero, and  (\ref{eqn_KKT2}) and (\ref{eqn_KKT3}) represent the complementary slackness. Based on the previous KKT conditions, we can recognize the interiors, boundaries, and outliers, as shown in Table \ref{tab_A1}. 
\begin{table}[]
\footnotesize
\centering
\vspace{-4mm}
\caption{Classification of samples}
\vspace{-2mm}
\begin{tabular}{ccc}
\hline
\textbf{Dual variables of samples} & \textbf{Positions} & \textbf{Categories} \\ \hline
\makecell*[c]{$\alpha_n=0$\\$\beta_n=1/(N\epsilon)$}  &$\Vert \bm \phi(\bm \xi^{(n)}) - \bm o \Vert^2 < R^2$& Interiors           \\\hline
\makecell*[c]{$0<\alpha_n<1/(N\epsilon)$\\$0<\beta_n<1/(N\epsilon)$}& $\Vert \bm \phi(\bm \xi^{(n)}) - \bm o \Vert^2 =R^2$ & Boundaries \\\hline
\makecell*[c]{$\alpha_n=1/(N\epsilon)$\\$\beta_n=0$} &   $\Vert \bm \phi(\bm \xi^{(n)}) - \bm o \Vert^2 >R^2$ &   Outliers\\\hline
\end{tabular} \label{tab_A1}
\vspace{-5mm}
\end{table}
We refer to the samples on the boundary as boundary support vectors, and regard outliers and boundary support vectors as support vectors. The sets of support vectors and boundary support vectors are defined based on Table \ref{tab_A1}:
\begin{align}
&\text{SV} = \{n|\alpha_n>0, \  \forall n \in \mathcal{N}\}, \label{eqn_SV}\\
&\text{BSV} = \{n|0<\alpha_n<1/(N\epsilon), \  \forall n \in \mathcal{N} \}. \label{eqn_BSV}
\end{align}
Note that the solutions of $\bm \alpha$ and $\bm \beta$ can be obtained by solving the dual problem of $\textbf{P-SVC}$, as follows
\begin{align}
&\min_{\bm \alpha} \sum_{n \in \mathcal{N}} \sum_{m \in \mathcal{N}} \alpha_n \alpha_m K(\bm \xi^{(n)},\bm \xi^{(m)}) - \sum_{n \in \mathcal{N}} K(\bm \xi^{(n)},\bm \xi^{(m)}), \tag{$\textbf{D-SVC}$}\\
&\begin{array}{r@{\quad}r@{}l@{\quad}l}
\text{s.t.} && 0 \geq \alpha_n \geq \frac{1}{N\epsilon}, \forall n \in \mathcal{N},
\end{array} \\
&\quad \quad \quad\sum_{n \in \mathcal{N}}\alpha_n=1, 
\end{align}
where $K(\bm \xi^{(n)},\bm \xi^{(m)})=\bm \phi(\bm \xi^{(n)})^\intercal \bm \phi(\bm \xi^{(m)})$ is the kernel function. This paper employs the weighted generalized intersection kernel \cite{shang2017data}   as our kernel function,
\begin{align}
K(\bm \xi^\text{(1)}, \bm \xi^\text{(2)}) = \sum_{d \in \mathcal{D}} l_d - \Vert\bm W (\bm \xi^\text{(1)}-\bm \xi^\text{(2)}) \Vert_1, \label{eqn_kernel}
\end{align}
where $\mathcal{D}=\{1,2,\cdots, D\}$ is the dimension index set of $\bm \xi^\text{(1)}$ or $\bm \xi^\text{(2)}$;  $l_d=\xi_{d,\text{max}}-\xi_{d,\text{min}}$, where  $\xi_{d,\text{max}}$ and $\xi_{d,\text{min}}$ are the upper and lower bounds, respectively, of $\bm \xi$ in the $d$-th dimension;  and $\bm W = \bm \Sigma^{-1/2}$, where $\bm \Sigma$ is the covariance matrix of the historical samples. Note that  $K(\bm \xi,\bm \xi)=\sum_{d \in \mathcal{D}} l_d$ is a constant. 

\begin{proposition} \label{prop_1}
The OC-SVC-based uncertainty set, i.e., $\mathcal{U}=\Vert \bm \phi(\bm \xi^{(n)}) - \bm o \Vert^2 \leq R^2$, can be expressed as a polyhedron in the original space of uncertainty $\bm \xi$, as follows:
\begin{align}
\mathcal{U}=\left\{
\bm \xi\left|
\begin{aligned}
&\exists \bm \upsilon_n \in \mathbb{R}^D, \forall n \in \text{SV}, \ \text{s.t.} \\
&\sum_{n \in \text{SV}}\alpha_n \bm \upsilon_n^\intercal \bm 1 \leq \gamma, \\
&-\bm \upsilon_n \leq \bm W(\bm \xi-\bm\xi^{(n)})\leq \bm \upsilon_n, \forall n \in \text{SV},
\end{aligned} \right.
\right\}, \label{eqn_uncertainty_set2}
\end{align}
where $\bm \upsilon_n$ is an auxiliary variable.
Parameter $\gamma$ is equal to $\sum_{n \in \text{SV}}\alpha_n\Vert \bm W(\bm \xi^{(k)} - \bm \xi^{(n)})\Vert_1$ for any $k \in \text{BSV}$.
\end{proposition} 
\emph{Proof}: see Appendix \ref{app_1}.

As mentioned above, a desirable uncertainty set must satisfy the feasibility, optimality, and tractability requirements. We show that the OC-SVC-based uncertainty set can meet the feasibility requirement.
\begin{proposition} \label{prop_11}
The uncertainty set $\mathcal{U}$ defined in  (\ref{eqn_uncertainty_set2}) can cover at least $100 (1-\epsilon)\%$ of samples.
\end{proposition}
\emph{Proof:} From Table \ref{tab_A1}, we know that any outlier must have $\alpha_n=1/(N\epsilon)$. If the number of outliers  exceeds $N\epsilon$, then $\bm 1^\intercal \bm \alpha > 1$ ($\alpha_n \geq 0$ for all samples), which conflicts with   KKT condition (\ref{eqn_KKT1}). Hence the number of outliers must be no larger than $N\epsilon$, and  the proposed uncertainty set can cover at least $100 (1-\epsilon)\%$ of samples.

We verify the optimality performance of the proposed uncertainty set based on numerical experiments in section \ref{sec_case}.

Since $\mathcal{U}$ is a polyhedron, we can find a linear counterpart for it, as we next discuss. Hence the tractability requirement can also be satisfied.

\subsection{Linear robust counterpart} \label{sec_counterpart}
We develop a linear counterpart for the robust approximation (\ref{eqn_RO}) to make the whole problem tractable.
First, based on (\ref{eqn_h_RO}), the robust approximation (\ref{eqn_RO}) can be expressed as 
\begin{align}
&\max_{\bm \xi_t \in \mathcal{U}_t} \left\lbrace\max_{\forall m \in \mathcal{M}} \big((\bm H_{m,t} \bm \xi_{t})^\intercal \bm y_{t} - \beta_{m,t}(\bm y_{t})\big)\right\rbrace \leq 0,\notag\\
&\Leftrightarrow \max_{\bm \xi_t \in \mathcal{U}_t}\big((\bm H_{m,t} \bm \xi_t)^\intercal \bm y_{t} \big)\leq \beta_{m,t}(\bm y_{t}),
\forall m \in \mathcal{M}.\label{eqn_RO_2}
\end{align}
Based on the following proposition, we can find a linear counterpart for the robust constraint (\ref{eqn_RO_2}).
\begin{proposition} \label{prop_2}
If the uncertainty set is defined by  (\ref{eqn_uncertainty_set2}), then  (\ref{eqn_RO_2}) can be  reformulated as linear constraints,
\begin{align}
\begin{cases}
\sum_{n \in \text{SV}_t}(\bm \mu_{n,m,t} - \bm \rho_{n,m,t})^\intercal \bm W_t \bm \xi_t^{(n)} + \pi_{m,t} {\gamma}_{t} \leq \beta_{m,t},\\
\sum_{n \in \text{SV}_t}\bm W_t (\bm \rho_{n,m,t} - \bm \mu_{n,m,t}) + \bm H_{m,t}^\intercal \bm y_{t} = \bm 0,\\
\bm \rho_{n,m,t} + \bm \mu_{n,m,t} = \pi_{m,t} \alpha_{n,t} \bm 1,\quad \pi_{m,t}\geq 0, \\
\bm \mu_{n,m,t},\bm \rho_{n,m,t} \in \mathbb{R}^D_+,\forall n \in \text{\rm SV}_t, \forall m \in \mathcal{M},\forall t \in \mathcal{T},
\end{cases} \label{eqn_convex}
\end{align}
where $\bm \rho_{n,m,t}$, $\bm \mu_{n,m,t}$, and $\pi_{m,t}$ are Lagrange multipliers for the $m$-th constraint at time $t$ in  (\ref{eqn_RO_2}).
\end{proposition}
\emph{Proof}: see Appendix \ref{app_2}.

\subsection{Summary}
Based on the previous linear counterpart, we can reformulate problem \textbf{P1} as
\begin{align} 
&\min_{\bm y_t, \forall t \in \mathcal{T}} \quad \sum_{t \in \mathcal{T}} \mathbb{E} \left(\sum_{t \in \mathcal{T}} EC_t \right)  \tag{$\textbf{P2}$},\\
&\begin{array}{r@{\quad}r@{}l@{\quad}l}
\text{s.t.} &\text{Eqs. } &\text{(\ref{eqn_thermal})-(\ref{eqn_comfort}), (\ref{eqn_EC})-(\ref{eqn_netload2}), (\ref{eqn_p})-(\ref{eqn_uncertain_DG}), (\ref{eqn_SOCP}), and (\ref{eqn_convex})}, 
\end{array} \notag
\end{align}

\begin{figure}
	\centering
	 \vspace{-4mm}
	\includegraphics[width=0.8\columnwidth]{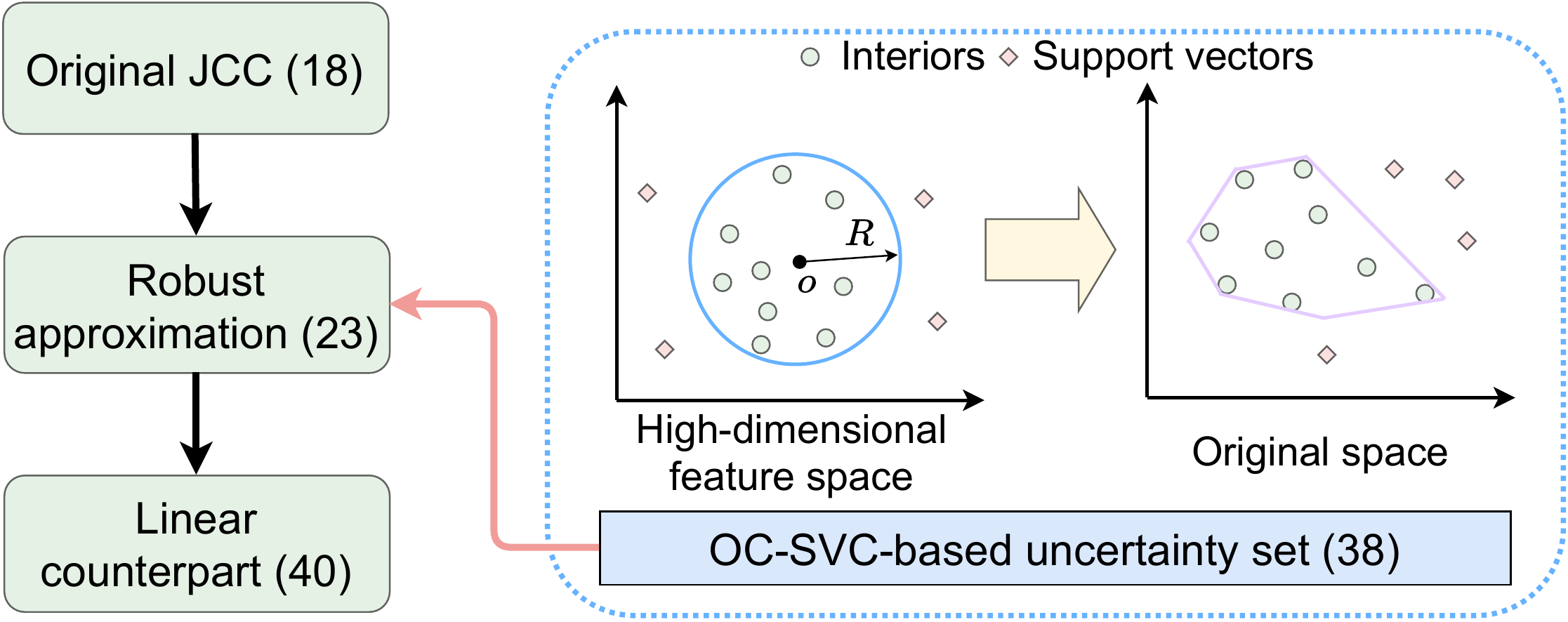}\vspace{-2mm}
	\caption{The whole procedure for converting JCCs into linear constraints.
	}
	\label{fig_procedure}
\end{figure}

Fig. \ref{fig_procedure} summarizes the procedure to convert intractable JCCs to tractable linear constraints. We  develop a robust optimization-based approximation (\ref{eqn_RO}) for the original JCC (\ref{eqn_JCC}), and solve the dual problem \textbf{D-SVC} to find the support vectors and construct   OC-SVC-based uncertainty set (\ref{eqn_uncertainty_set2}). We implement the linear counterpart based on  (\ref{eqn_convex}) to replace the original intractable JCC, and solve problem \textbf{P2} to obtain the optimal schedule of HVAC systems and DRG.

\section{Case study} \label{sec_case}
\subsection{Simulation set up}
\begin{figure}
	\subfigbottomskip=-6pt
	\subfigcapskip=-4pt
		\vspace{-4mm}
	\centering
	\subfigure[]{\includegraphics[width=0.49\columnwidth]{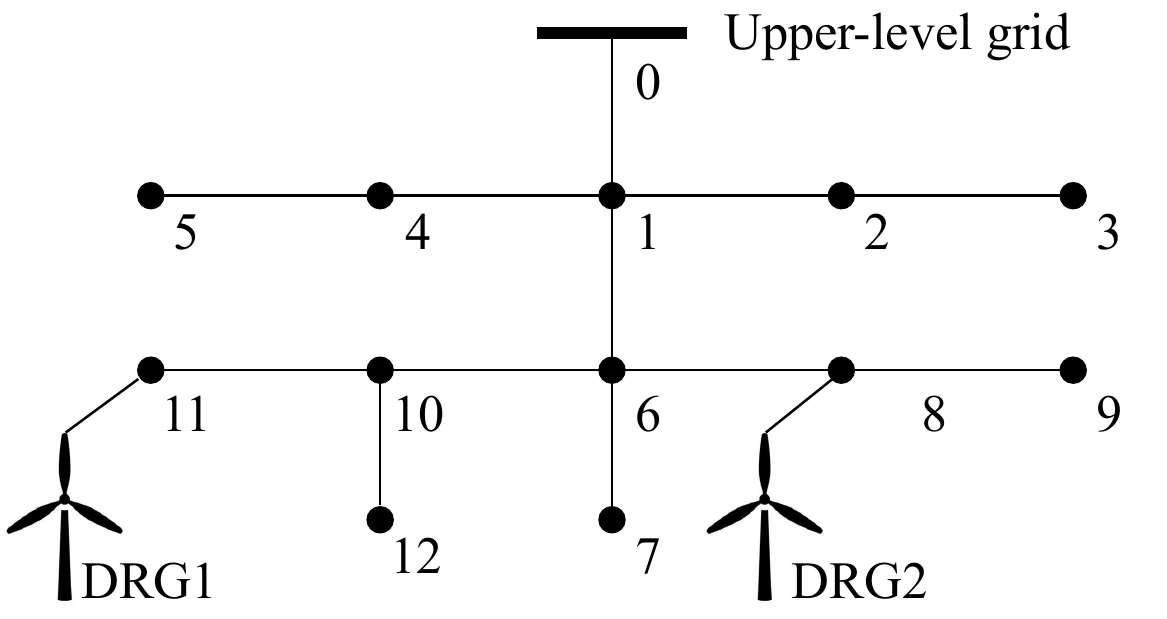}}
	\subfigure[]{\includegraphics[width=0.49\columnwidth]{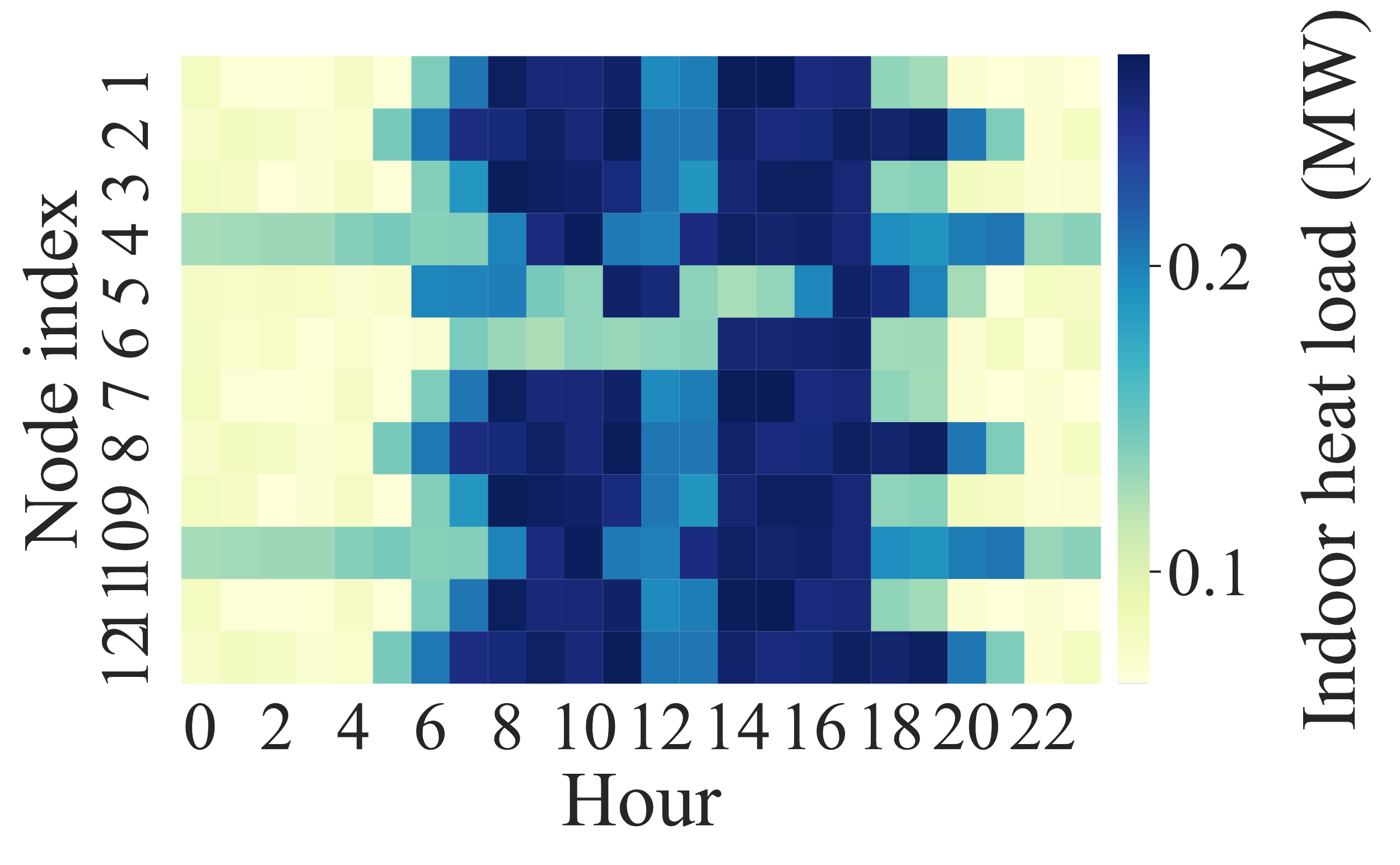}}
	\subfigure[]{\includegraphics[width=0.49\columnwidth]{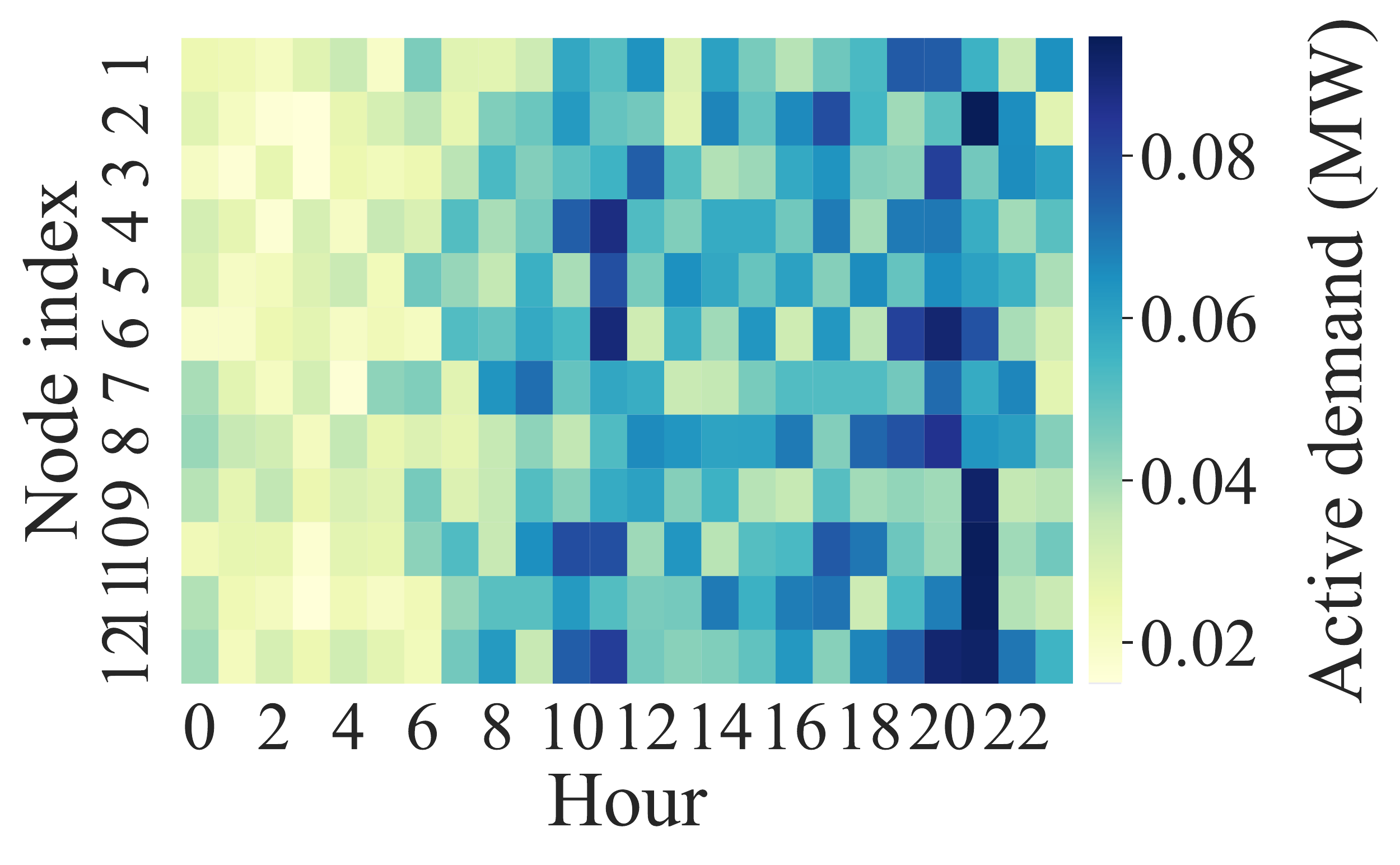}}
	\subfigure[]{\includegraphics[width=0.49\columnwidth]{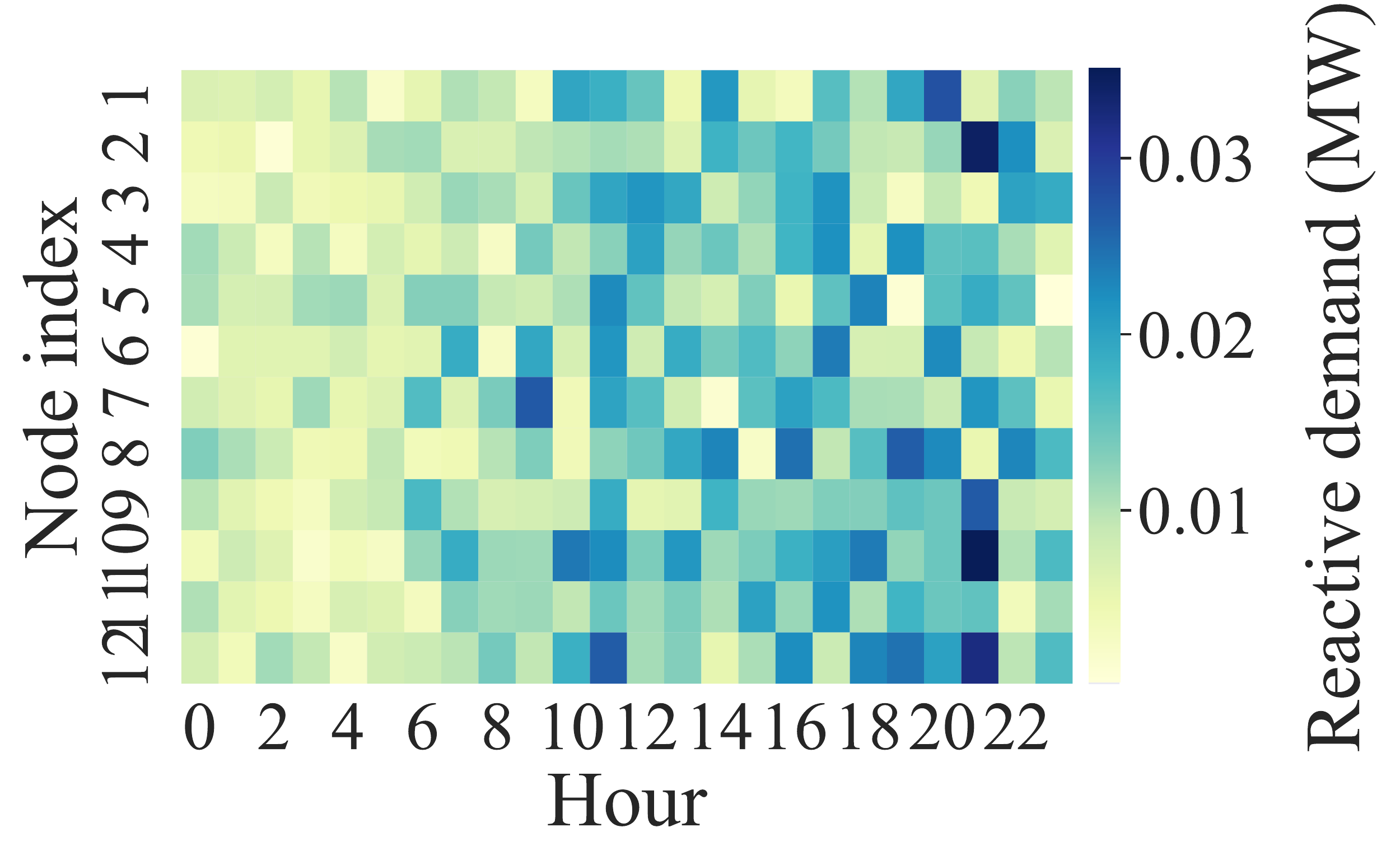}}
	\vspace{-4mm}
	\caption{{(a) The 13-Bus test system, (b) indoor heat loads, (c), base active power demands, and (d) base reactive power demands. In (a), symbols ``DRG1" and ``DRG2" represent two distributed renewable generators. }}
	\label{fig_parameters}
	\vspace{-4mm}
\end{figure}

\begin{figure}
	\subfigbottomskip=-6pt
	\subfigcapskip=-4pt
		\vspace{-4mm}
	\centering
	\subfigure[]{\includegraphics[width=0.49\columnwidth]{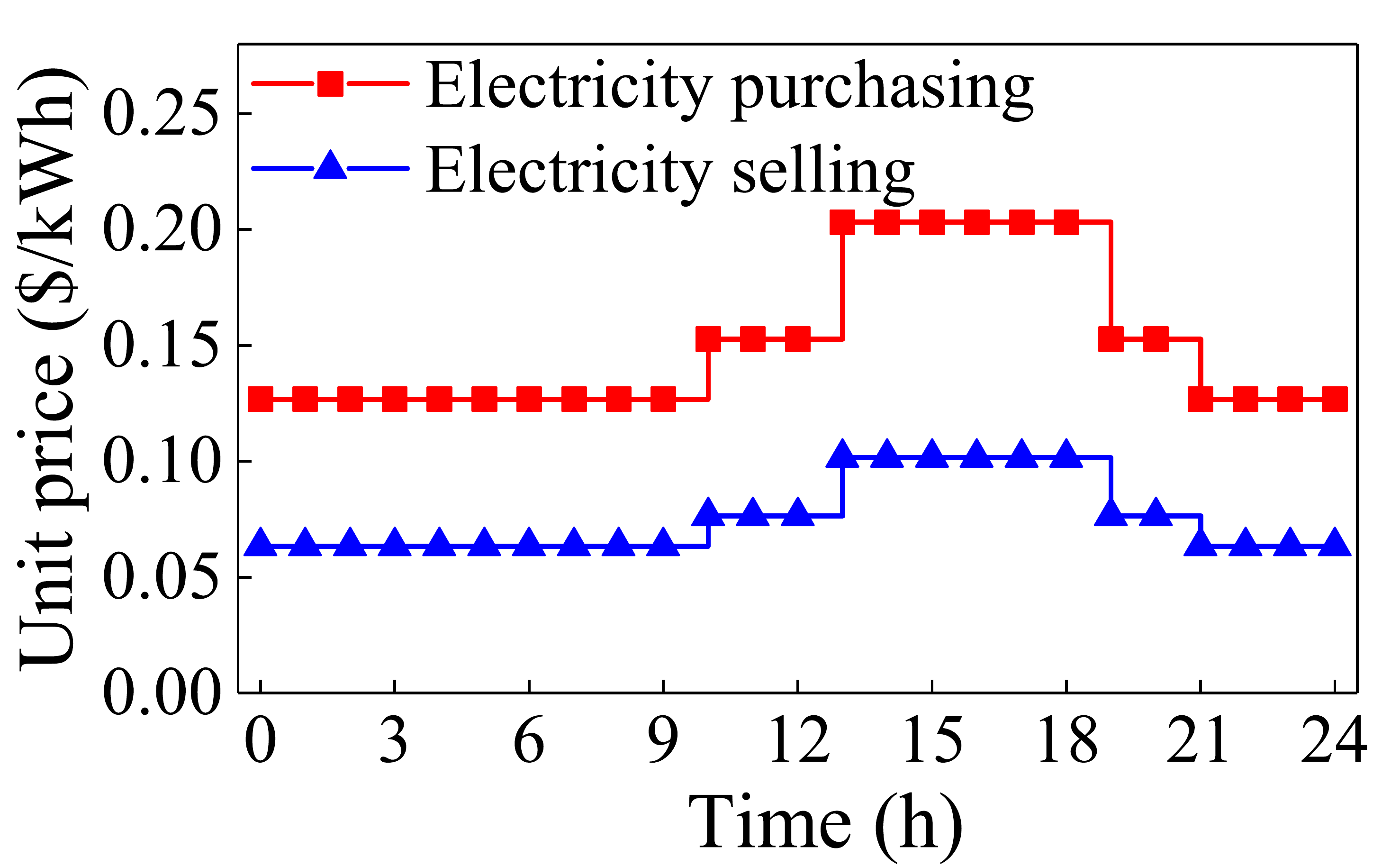}}
	\subfigure[]{\includegraphics[width=0.49\columnwidth]{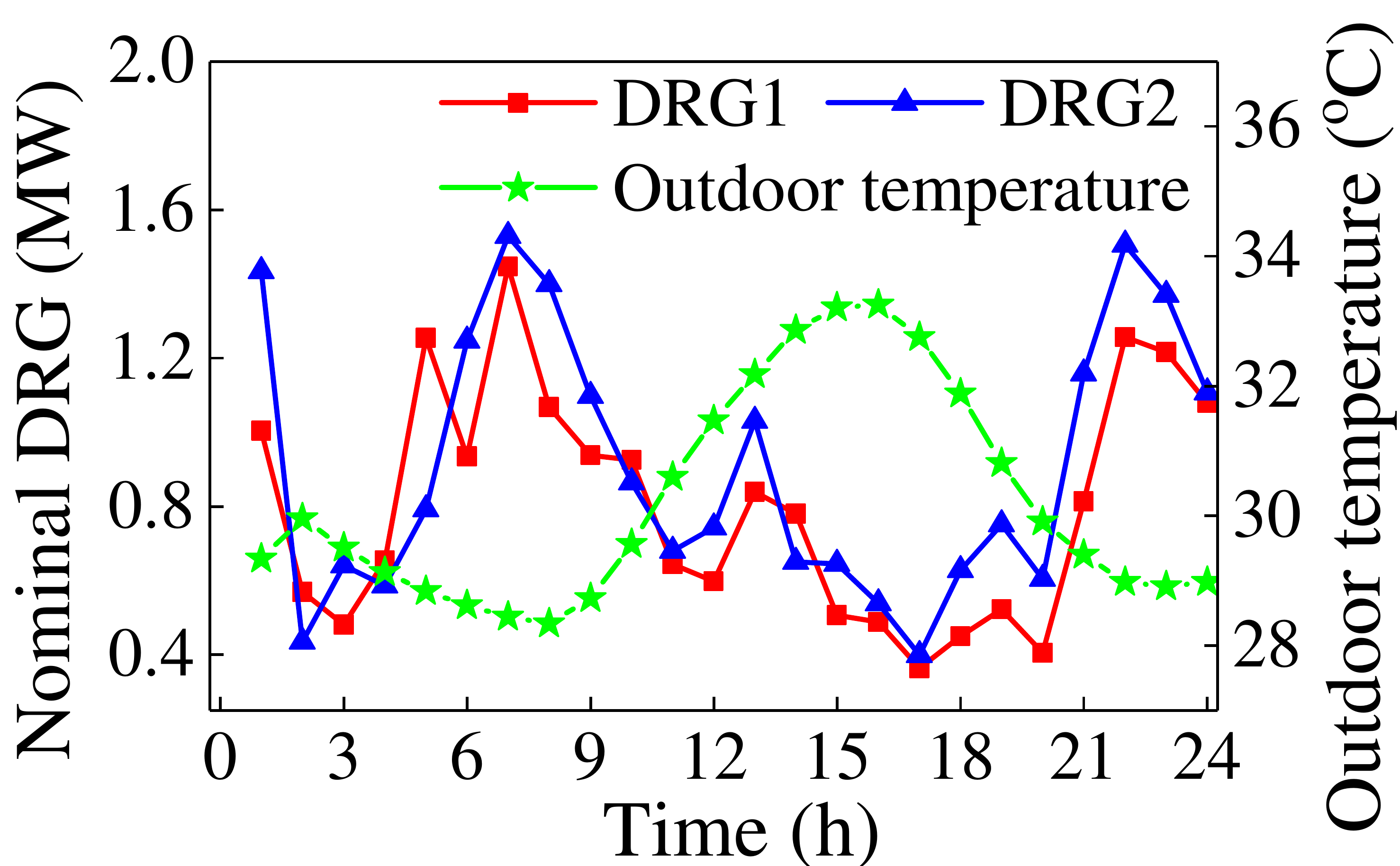}}
	\vspace{-4mm}
	\caption{{(a) Unit prices of electricity purchasing and selling and (b) nominal outputs of DRG and outdoor temperature.}}
	\label{fig_parameters2}
\end{figure}

{We describe a case study based on the IEEE 13-bus system with wind turbines DRG1 and DRG2 to provide DRG, as shown in Fig. \ref{fig_parameters}(a). The network parameters (e.g., branch resistance and reactance) can be found on the IEEE  website \cite{Resources}. The indoor heat loads and base power demands (i.e.,   demands except HVAC loads) in different nodes are illustrated in Fig. \ref{fig_parameters}(b)--(d). The unit prices for purchasing and selling electricity are shown in Fig. \ref{fig_parameters2}(a), and  Fig. \ref{fig_parameters2}(b) shows the nominal outputs of DRG and outdoor temperature. The optimization horizon is set as 24 hours, and the time interval is one hour. Other parameters can be found in Table \ref{tab_parameter}.} Because the forecasting errors of wind generation can be variant, we implement \textbf{Cases 1--3} with samples generated by Beta, Weibull, and Gaussian distributions, respectively, to simulate the historical data. The generated data has been uploaded on \cite{samples2021}. 

All numerical experiments are tested on an Intel 8700 3.20-GHz CPU with 16 GB memory. GUROBI and CVXPY are employed to solve the optimization problem. 
\begin{table}
	\small
	\centering
	\vspace{-4mm}
	\caption{{Parameters in case study}}
	\vspace{-2mm}
	\begin{footnotesize}
	\begin{tabular}{cccc}
		\hline
		\rule{0pt}{11pt}		
		Parameters & Value &Parameters & Value\\
		\hline
		\rule{0pt}{10pt}
		$C_i$ &  1 MWh/{\textcelsius} & $\overline{p}_i$ & 0.5MW \\
		$R_i$ &  20 {\textcelsius}/MW &  $\phi_i$ & 0.98 \\
		$\text{COP}_i$ &  3.6 &  $U_{i,\text{min}}$ & 0.95 p.u. \\
		 $\underline \theta_i$ & 24{\textcelsius} &$U_{i,\text{max}}$ & 1.05 p.u. \\
		$\overline \theta_i$& 28{\textcelsius} & $S_{ij,t}$ & 2 MW \\
		\hline
	\end{tabular}\label{tab_parameter}
	\end{footnotesize}
\end{table}
\subsection{Benchmarks}
We introduce three models as   benchmarks:
{
\begin{enumerate}
    \item \textbf{B1(SA+Box)}: Scenario approach combined with a box uncertainty set \cite{8060613, 8626040};
	\item \textbf{B2(SA+CH)}: Scenario approach combined with a convex hull uncertainty set \cite{8706676};
	\item \textbf{B3(Bonferroni)}: Bonferroni approximation to decompose   original JCC into multiple ICCs \cite{hassan2018optimal,odetayo2018chance}.
\end{enumerate}}
We also tested the performance of SAA. However, since it requires multiple constraints for every historical sample, the out-of-memory issue occurs, which implies a huge computational burden. {Parameters used in simulations are consistent among all methods, so their results can be compared.}


\subsection{Case 1: Beta distributed uncertainties}
\subsubsection{Shape of uncertainty sets}

\begin{figure}
		\vspace{-4mm}
	\subfigbottomskip=-4pt
	\subfigcapskip=-4pt
	\centering
	\subfigure[]{\includegraphics[width=0.49\columnwidth]{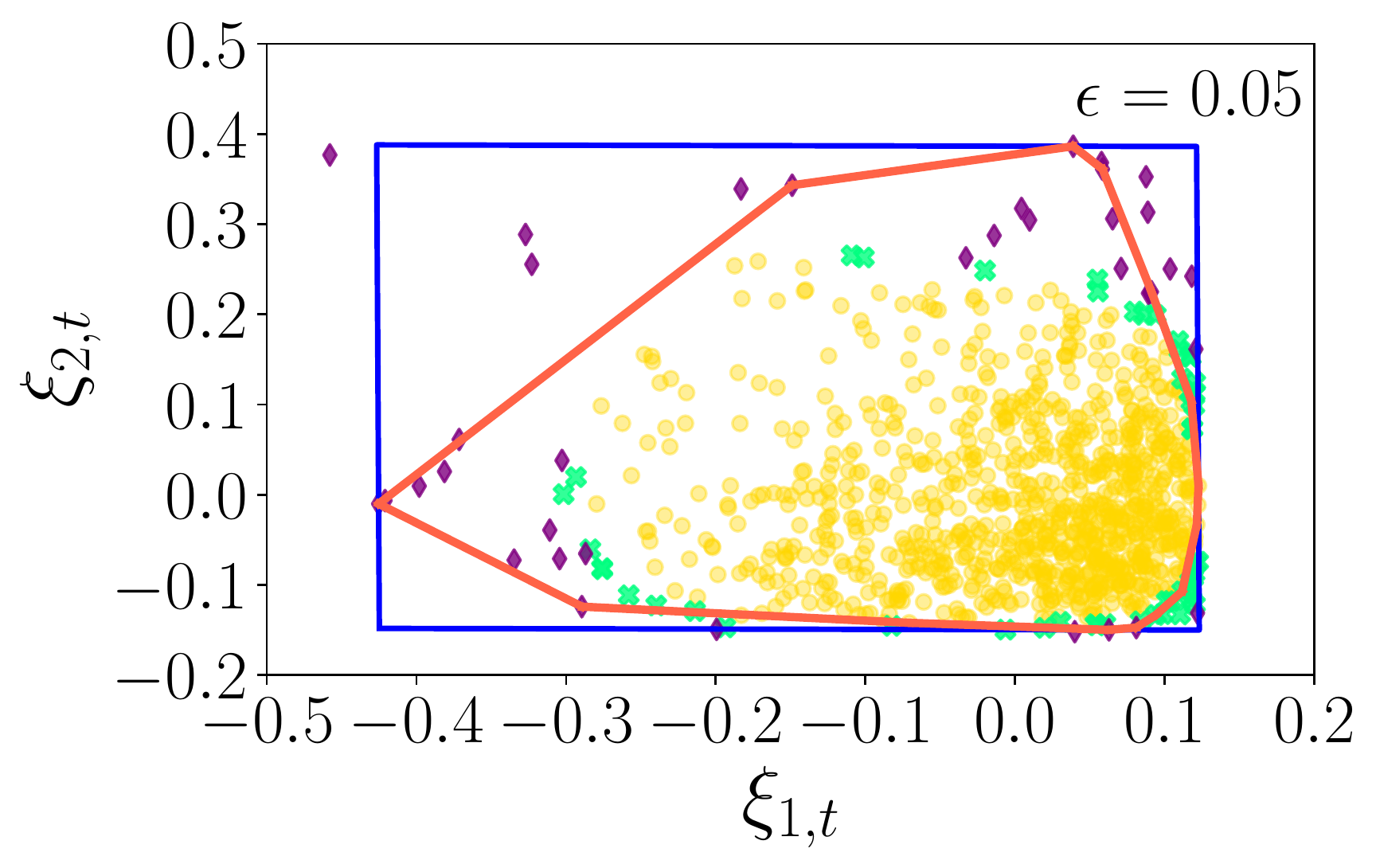}}
	\subfigure[]{\includegraphics[width=0.49\columnwidth]{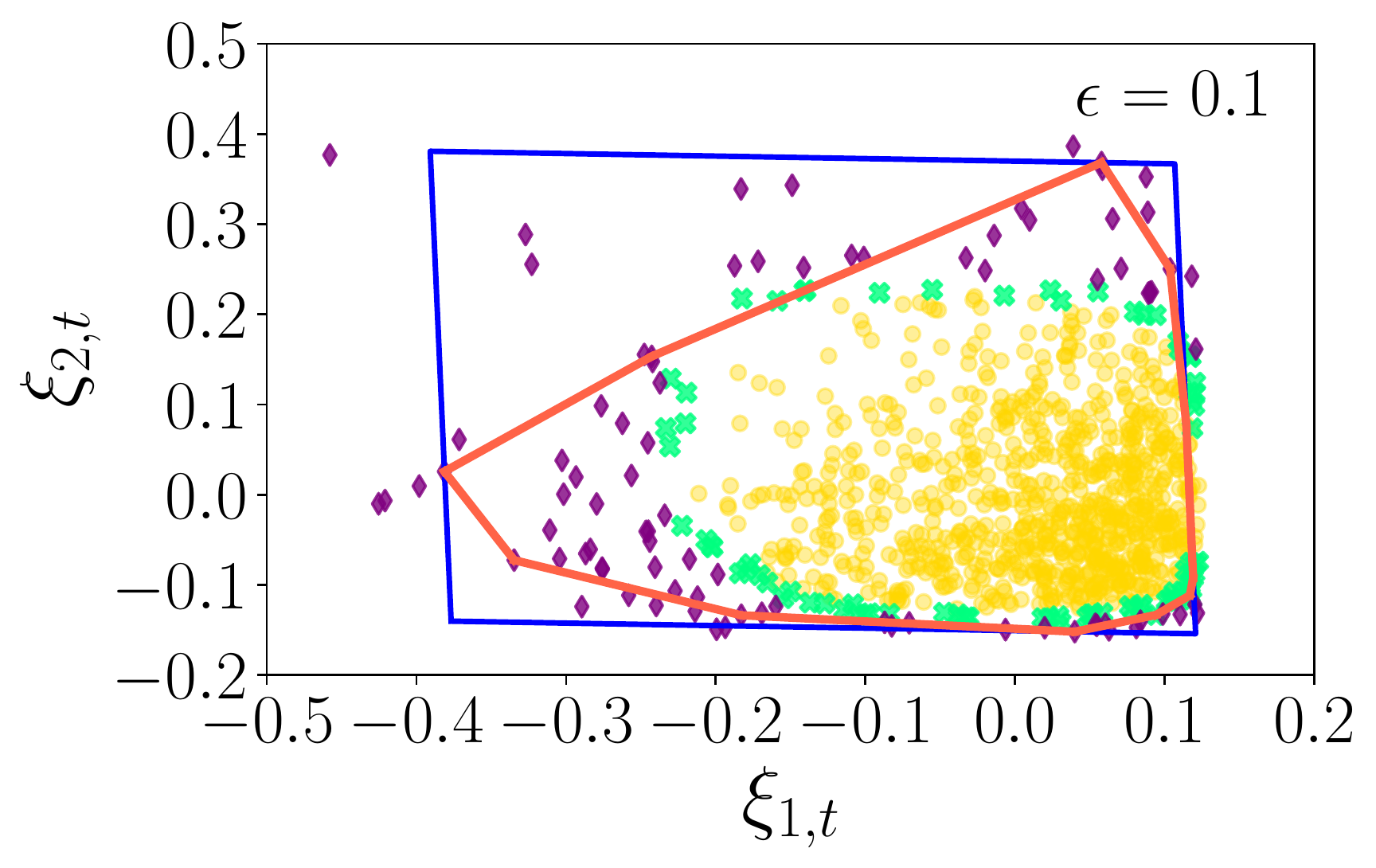}}
	\subfigure[]{\includegraphics[width=0.49\columnwidth]{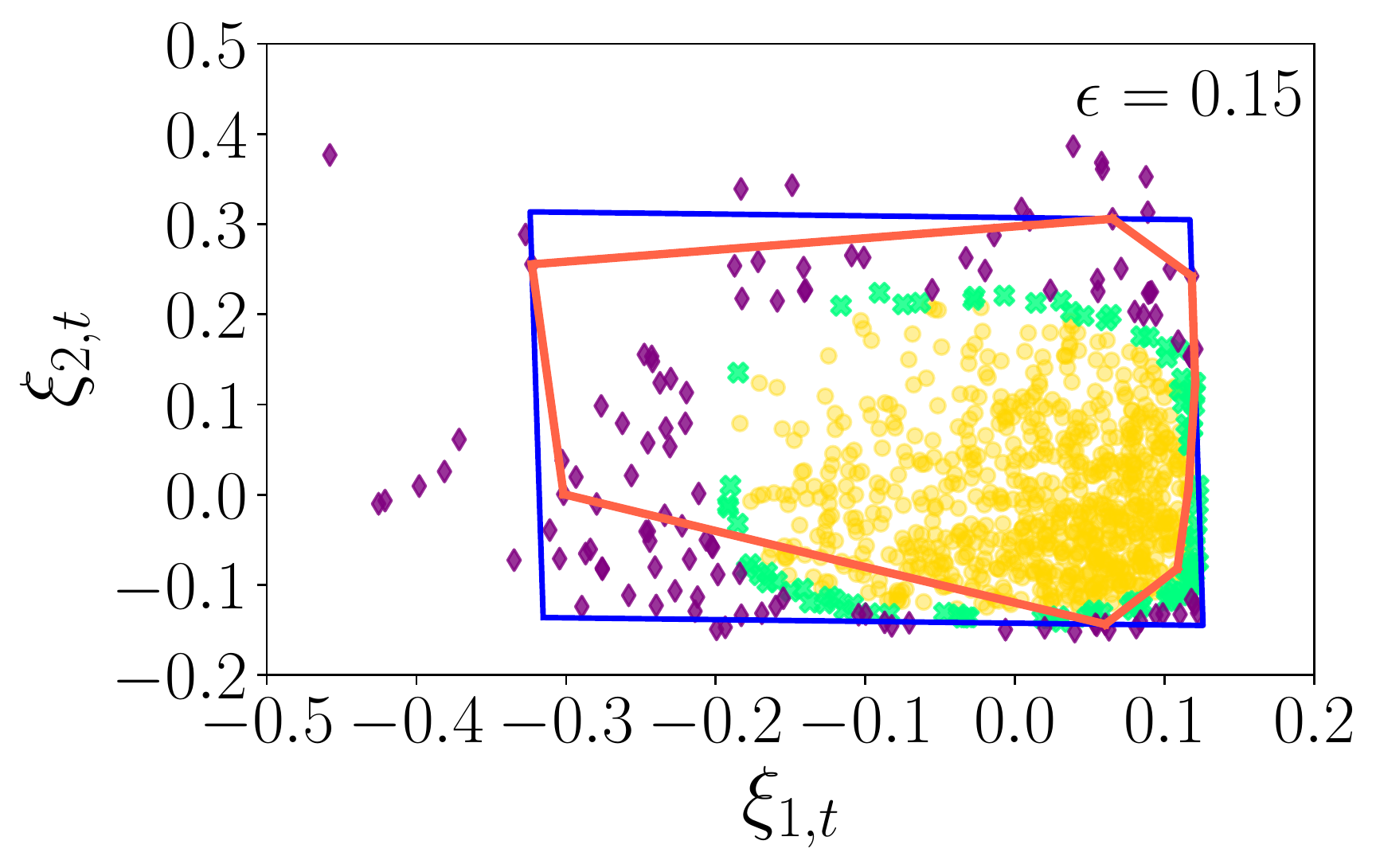}}
	\subfigure[]{\includegraphics[width=0.49\columnwidth]{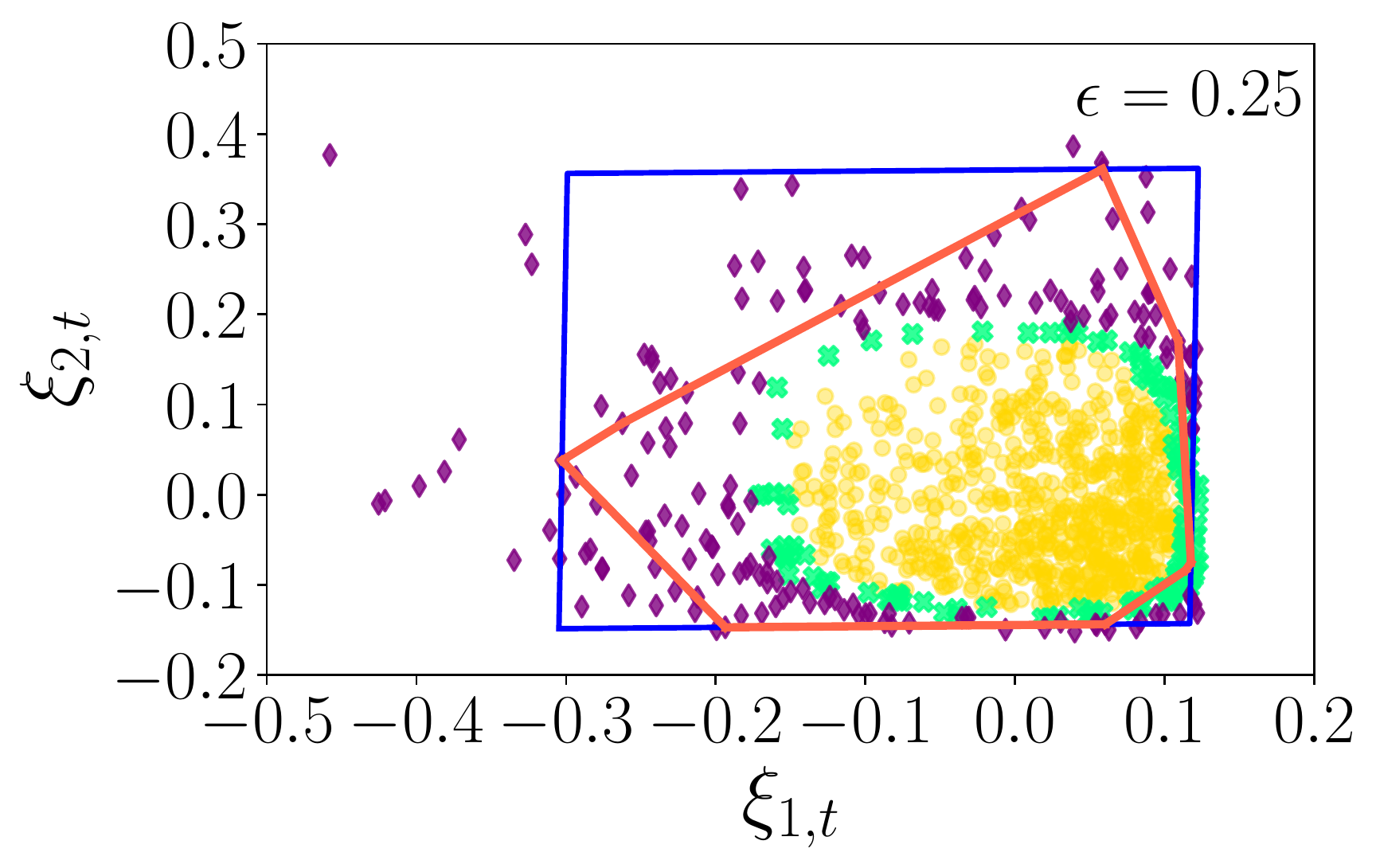}}
 	\caption{Uncertainty sets in \textbf{Case 1} (uncertainties follow Beta distribution) generated by different models at $t=6$: (a) $\epsilon=0.05$; (b) $\epsilon=0.10$; (c) $\epsilon=0.15$; (d) $\epsilon=0.25$. Blue rectangle and red polyhedron indicate uncertainty sets constructed by \textbf{B1(SA+Box)} and \textbf{B2(SA+CH)}, respectively.   SVC-based uncertainty set is formed by boundary support vectors (green points). Internal samples and outliers are marked as gold dots and purple diamonds, respectively.}
	\label{fig_uncertaintySet_case1}
	\vspace{-4mm}
\end{figure}

Fig. \ref{fig_uncertaintySet_case1} illustrates the shapes of uncertainty sets constructed by different models based on the samples at $t=6$ in \textbf{Case 1}. Since the scenarios are randomly drawn, some extreme samples are chosen. Hence, the uncertainty sets of   scenario approach-based models \textbf{B1} and \textbf{B2} must be large enough to cover these extreme samples, which {is} unnecessarily conservative. Unlike \textbf{B1} and \textbf{B2}, the proposed method introduces OC-SVC to decide which samples must be covered. By solving \textbf{P-SVC},   samples that are far from the majority of samples can be recognized as outliers. Compared to \textbf{B1} and \textbf{B2},   samples covered by the proposed uncertainty set are more concentrated, and the proposed uncertainty set (bounded by   green cross in Fig. \ref{fig_uncertaintySet_case1}) is much smaller. These results indicate that the proposed method can achieve better optimality than \textbf{B1} and \textbf{B2}.

\subsubsection{Optimality, feasibility and time-efficiency}

\begin{figure}
	\centering
	 			\vspace{-4mm}
	\includegraphics[width=1\columnwidth]{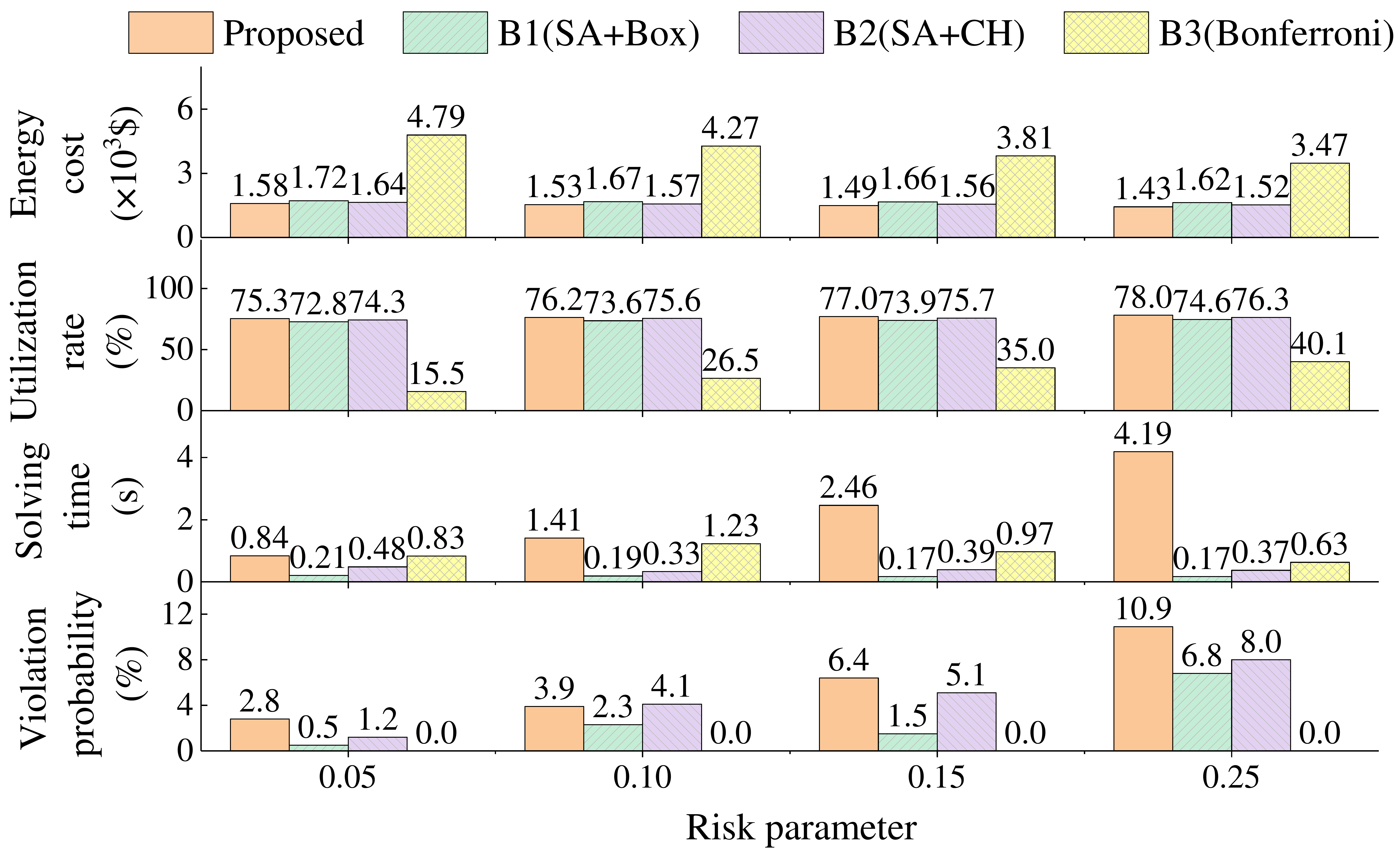}\vspace{-2mm}
	\caption{Results of energy costs, average utilization rates of DRG, solving times, and maximum violation probabilities obtained by different models in \textbf{Case 1} (Uncertainties follow the Beta distribution).
	}
	\label{fig_Beta}
	 		\vspace{-4mm}
\end{figure}

Fig. \ref{fig_Beta} summarizes the energy cost, average utilization rate of DRG, solution time, and maximum violation probability obtained by different models. {The energy cost of Bonferroni approximation \textbf{B3} is the worst among all models because the number of constraints in the original JCC (\ref{eqn_JCC}) is relatively large, and the risk parameter of each ICC is too small (less than 0.0015), leading to a   conservative solution. The proposed method achieves the lowest energy cost and highest utilization rate of DRG. The proposed OC-SVC-based uncertainty set can cover the historical samples more tightly than conventional box and convex hull uncertainty sets. Hence the solution is less conservative (i.e., the energy cost is smaller), and more DRG can be utilized in distribution networks; e.g., the average utilization rate of DRG obtained by the proposed model is around 2.5\% and 1.5\% higher than in \textbf{B1} and \textbf{B2}, respectively. These results confirm that the proposed model can better promote DRG utilization.}

According to  (\ref{eqn_convex}), the proposed model must introduce multiple dual variables for each support vector, so the computational burden is larger compared to the other models. Nevertheless, since the number of support vectors is much less than the total number of samples, {its computational efficiency is acceptable (in all cases, its maximum solution time is around 4s). Given that we focus on the daily schedule of HVAC systems (optimization horizon is 24 h), this is acceptable. The proposed method's solution time decreases with the risk parameter's decrease. For example, when $\epsilon=0.05$, it can complete the solution process in 1 s. Considering that the risk parameter is commonly maintained at a low level to guarantee the quality of service, this can meet the computational efficiency requirements in practice. }

{The violation probabilities of the original JCC (\ref{eqn_JCC}) obtained by all models, including the proposed one, are always lower than the given risk parameter.} For \textbf{B1}--\textbf{B3}, it has been pointed out \cite{geng2019data}  that they provide conservative approximations for JCCs. Eqs. (\ref{eqn_JCC_generic})--(\ref{eqn_RO}) imply that the proposed model is also an inner approximation. Hence all four models can guarantee feasible solutions.

\subsection{Case 2: Weibull distributed uncertainties}
\subsubsection{Shape of uncertainty sets}
\begin{figure}
		\vspace{-4mm}
	\subfigbottomskip=-4pt
	\subfigcapskip=-4pt
	\centering
	\subfigure[]{\includegraphics[width=0.49\columnwidth]{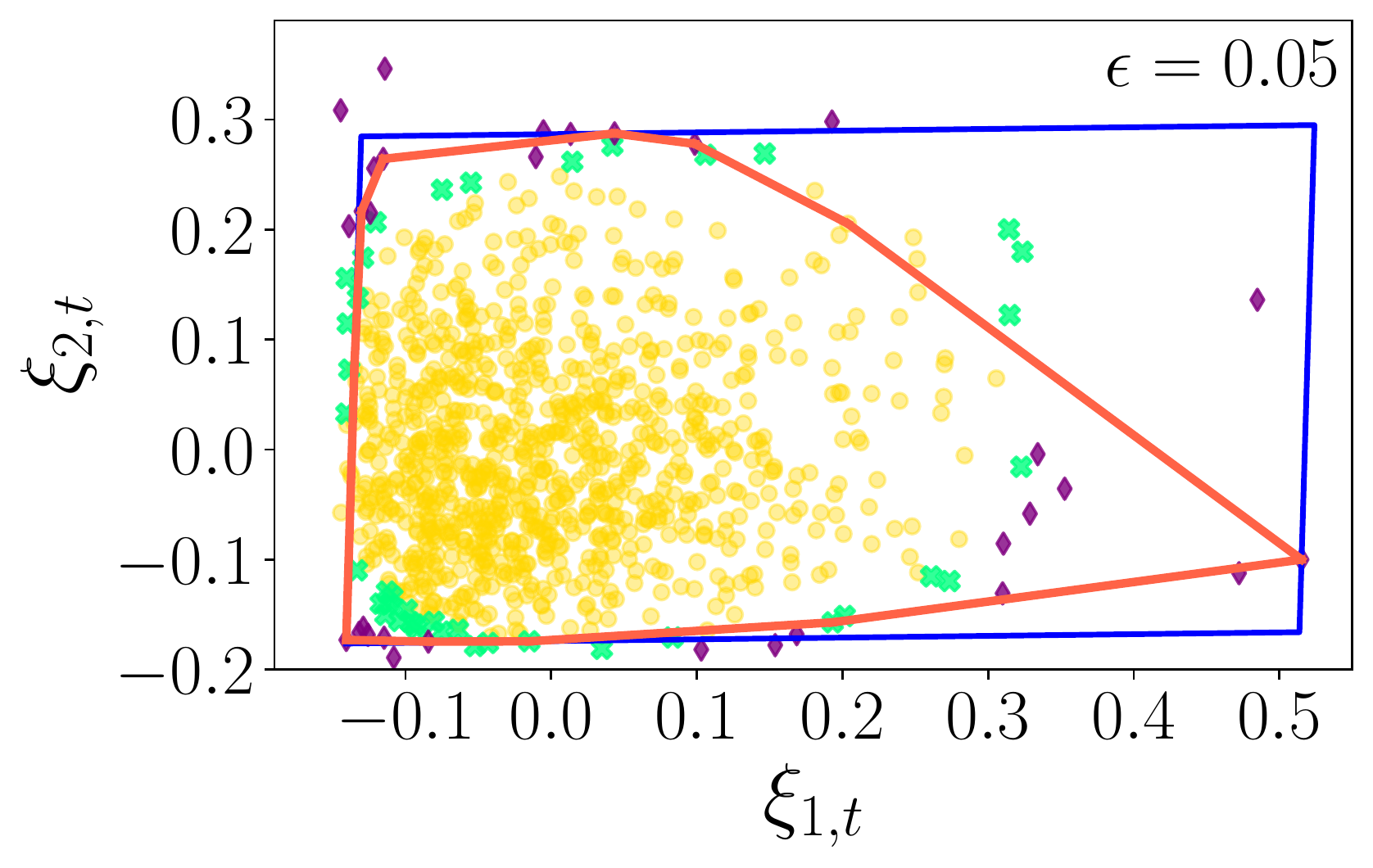}}
	\subfigure[]{\includegraphics[width=0.49\columnwidth]{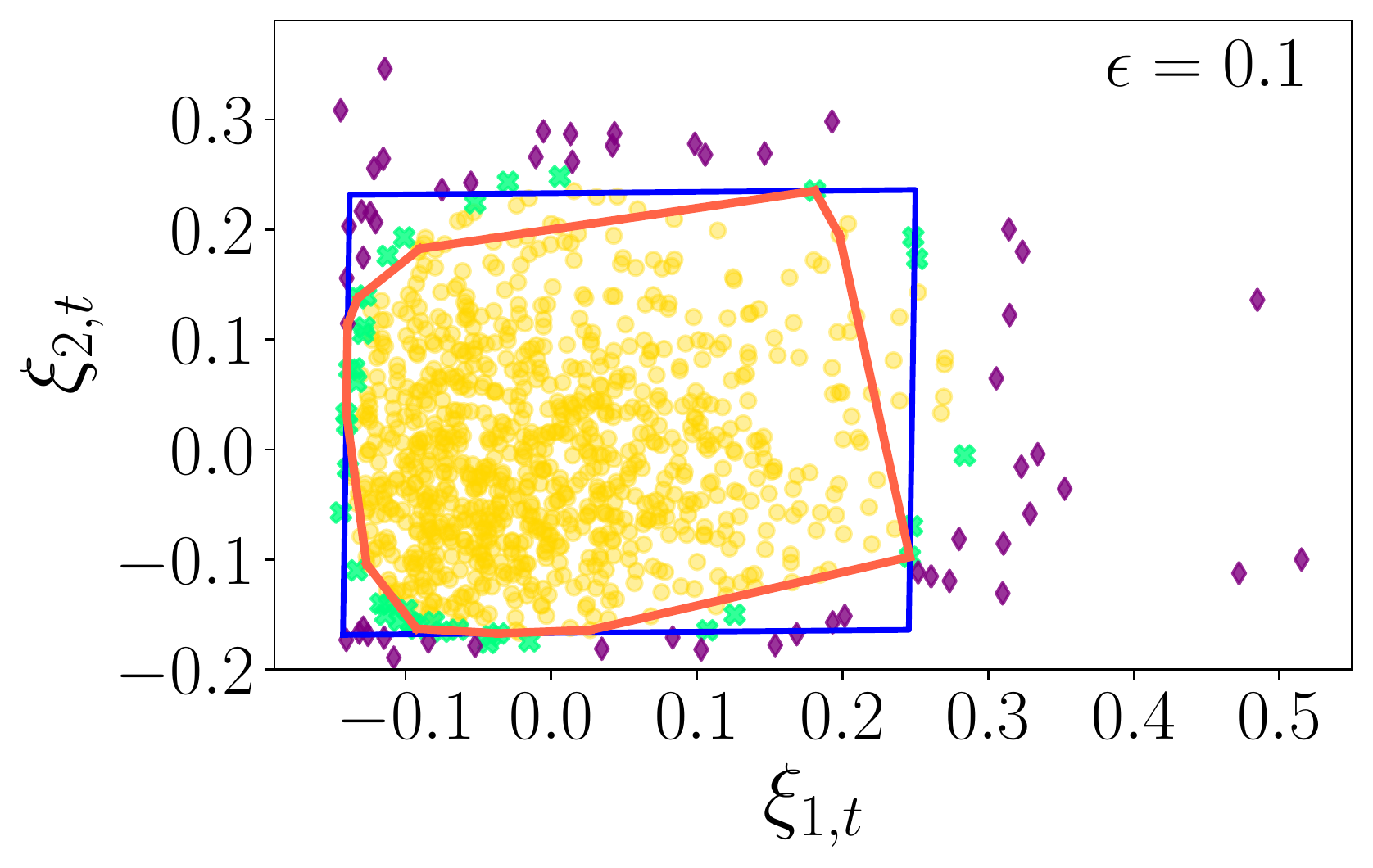}}
	\subfigure[]{\includegraphics[width=0.49\columnwidth]{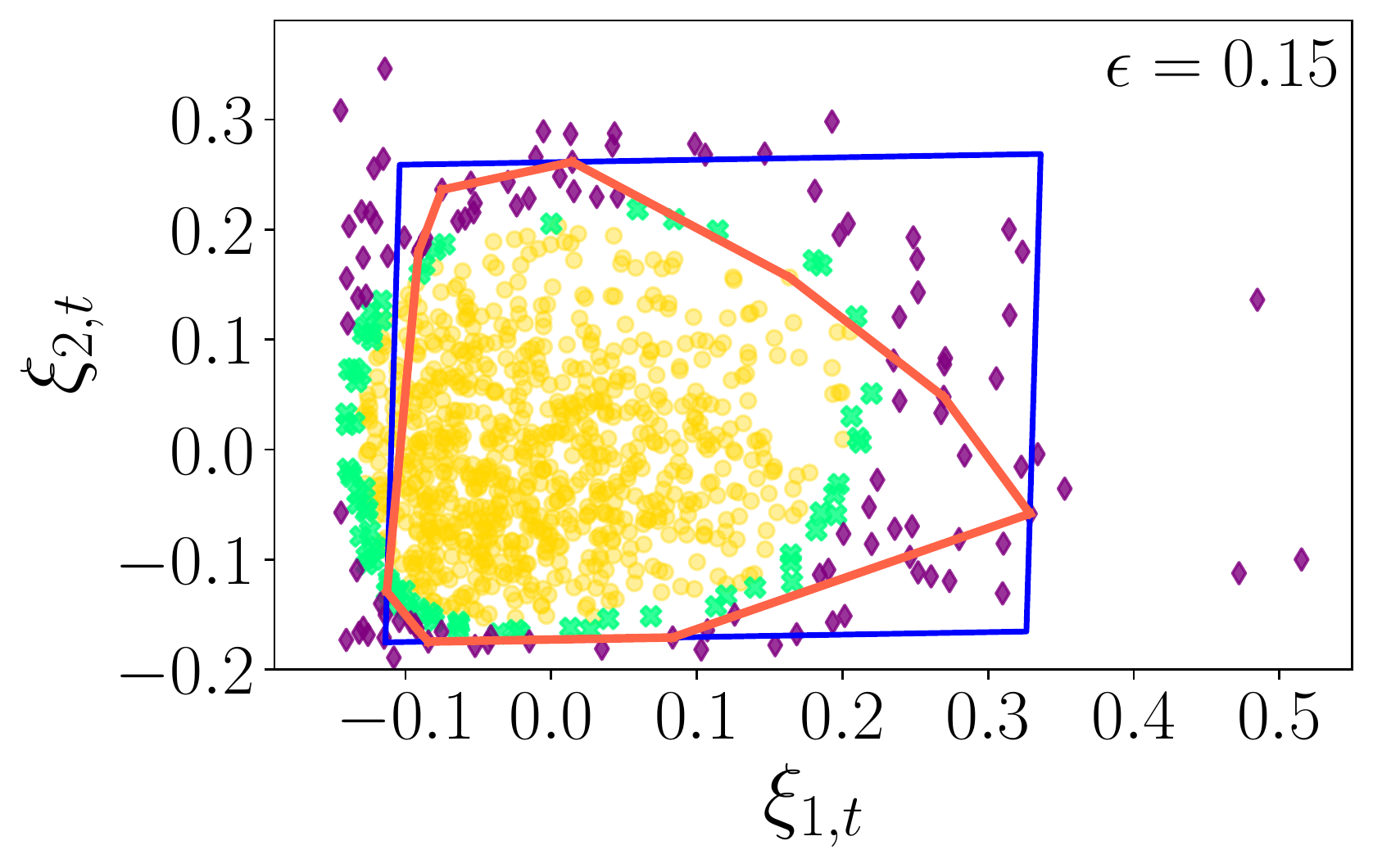}}
	\subfigure[]{\includegraphics[width=0.49\columnwidth]{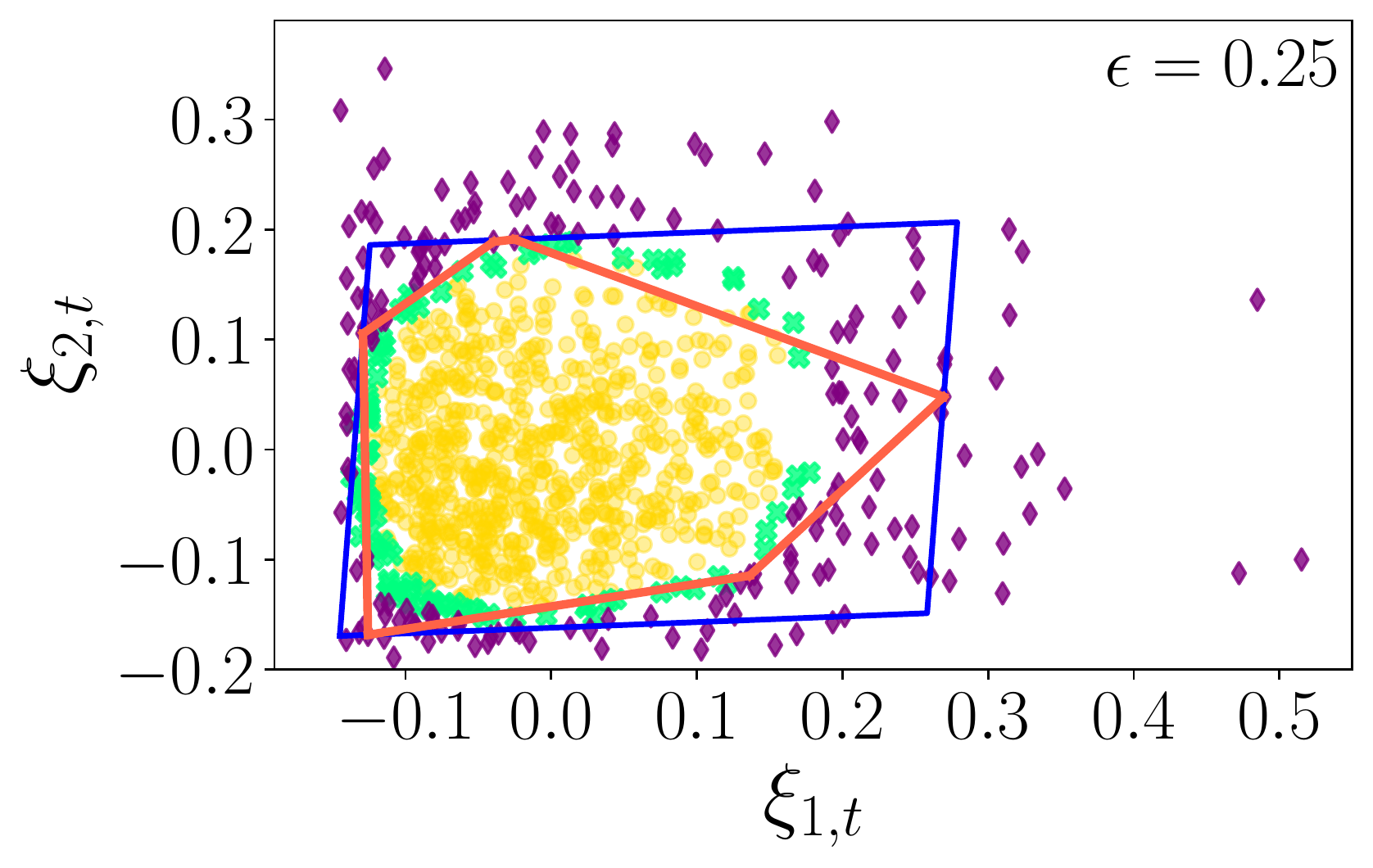}}
 	\caption{Uncertainty sets generated by different models in \textbf{Case 2} (uncertainties follow   Weibull distribution) at $t=6$: (a) $\epsilon=0.05$; (b) $\epsilon=0.10$; (c) $\epsilon=0.15$;   (d) $\epsilon=0.25$.}
	\label{fig_uncertaintySet_case2}
	\vspace{-4mm}
\end{figure}

Fig. \ref{fig_uncertaintySet_case2} shows the uncertainty sets established by different models in \textbf{Case 2}. Similar to \textbf{Case 1}, the proposed OC-SVC-based method can capture the characteristics of the Weibull distribution and tightly cover most samples. Therefore, the unnecessary space introduced by the proposed uncertainty set is smaller 
than those of scenario approaches \textbf{B1} and \textbf{B2}. 

\subsubsection{Optimality, feasibility and time-efficiency}
\begin{figure}
	\centering
	\includegraphics[width=1\columnwidth]{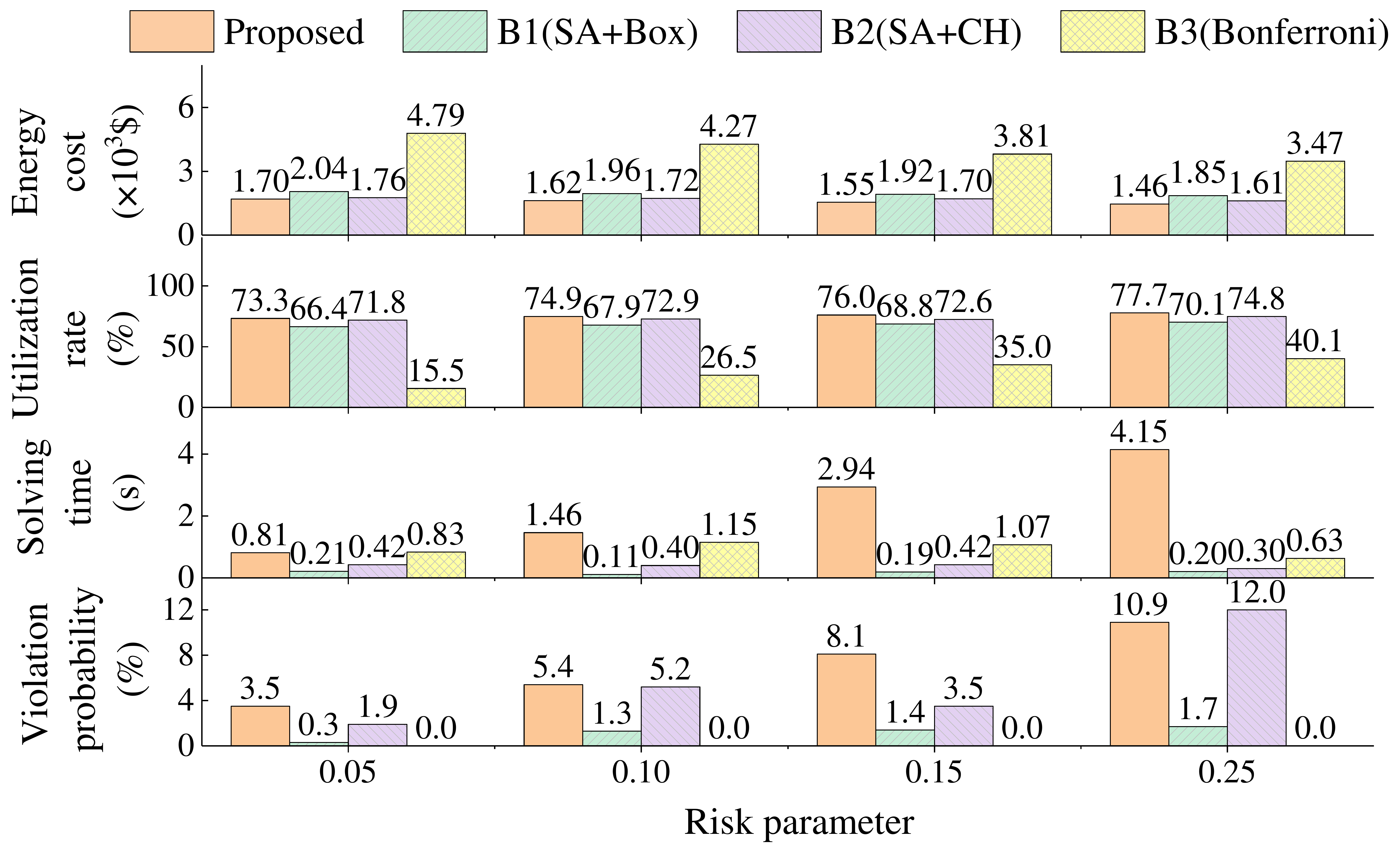}\vspace{-2mm}
	\caption{Results of energy costs, average utilization rates of DRG, solution times, and maximum violation probabilities obtained by different models in \textbf{Case 2} (uncertainties follow Weibull distribution).
	}
	\label{fig_Weibull}
	 		\vspace{-4mm}
\end{figure}

The results of different models in \textbf{Case 2} are listed in Fig. \ref{fig_Weibull}. Bonferroni approximation \textbf{B3} shows the worst energy efficiency because of the small risk parameter in each ICC. With the smallest volume of the OC-SVC-based uncertainty set, the proposed model achieves the highest energy efficiency and utilization rate of DRG with proper feasibility.

\subsection{Case 3: Normally distributed uncertainties}
\subsubsection{Shape of uncertainty sets}
\begin{figure}
		\vspace{-4mm}
	\subfigbottomskip=-4pt
	\subfigcapskip=-4pt
	\centering
	\subfigure[]{\includegraphics[width=0.49\columnwidth]{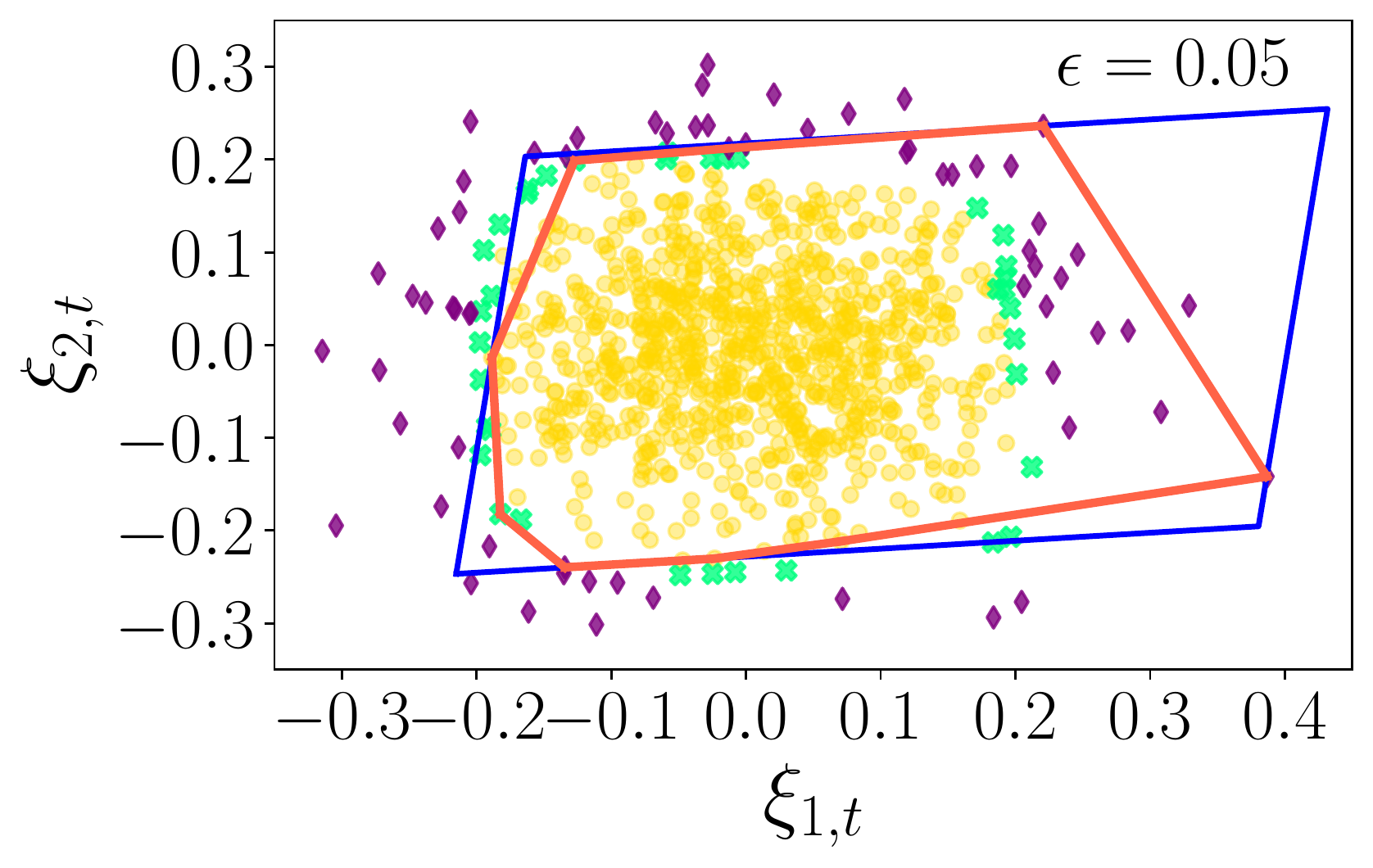}}
	\subfigure[]{\includegraphics[width=0.49\columnwidth]{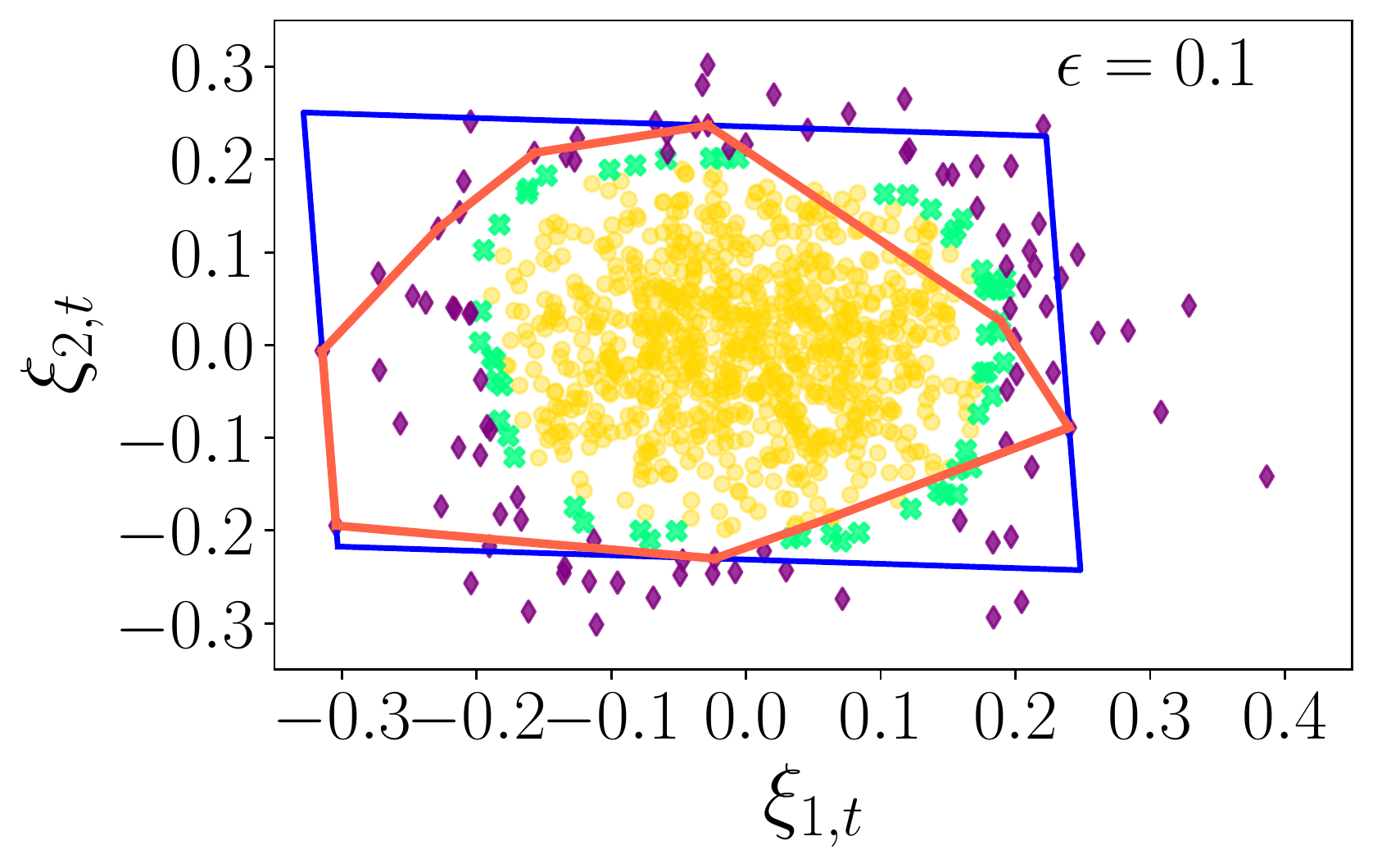}}
	\subfigure[]{\includegraphics[width=0.49\columnwidth]{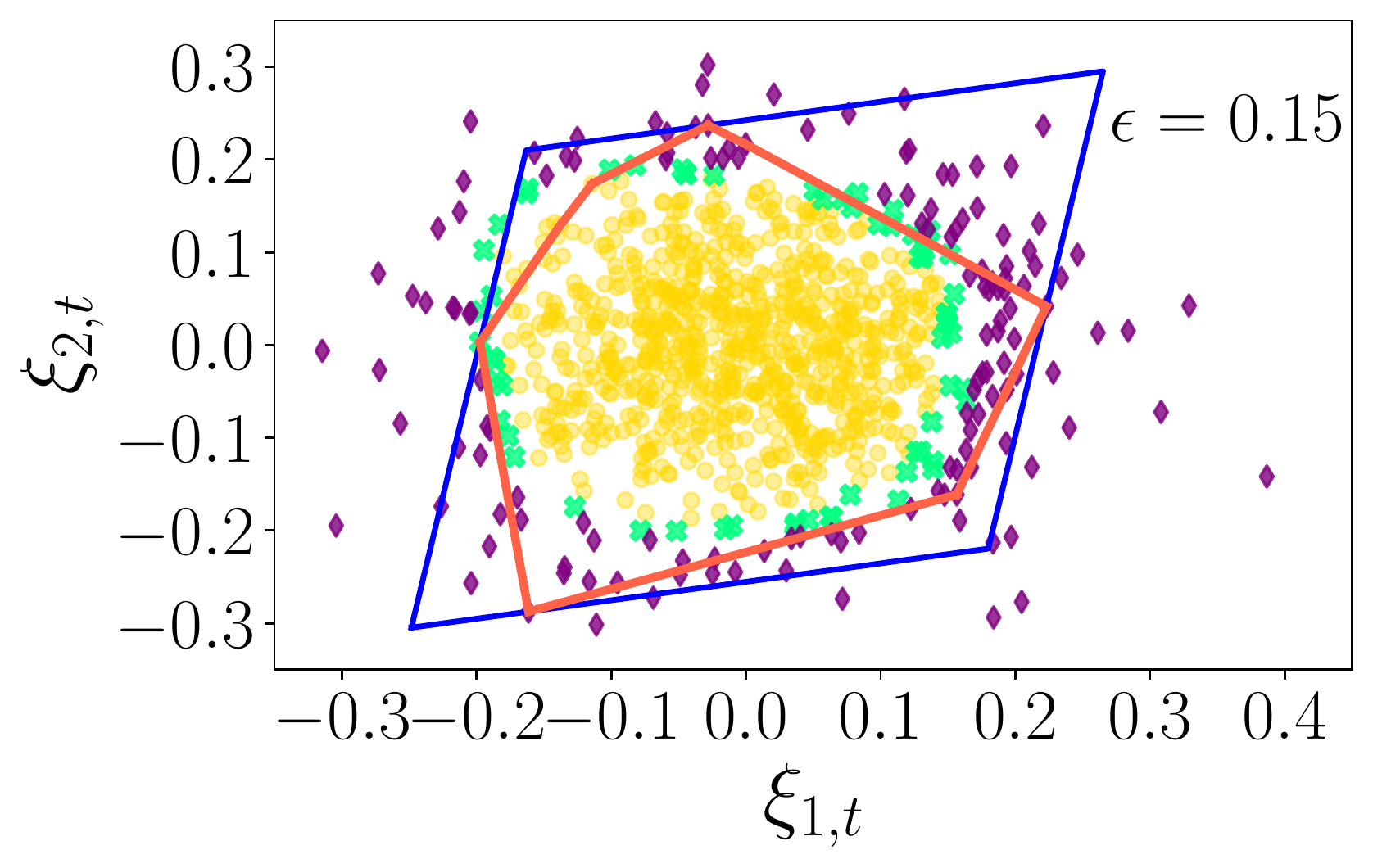}}
	\subfigure[]{\includegraphics[width=0.49\columnwidth]{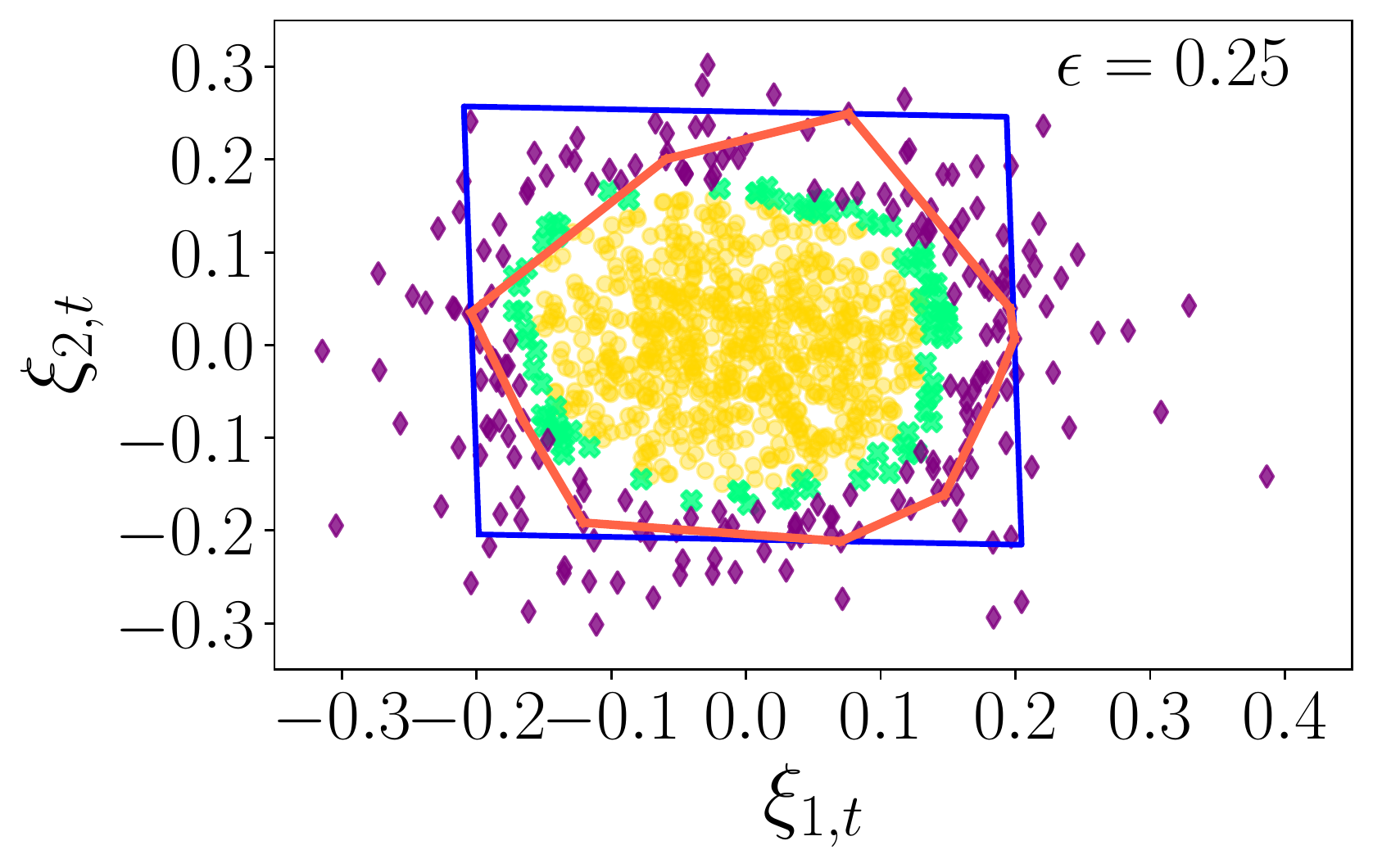}}
 	\caption{Uncertainty sets generated by different models in \textbf{Case 3} (uncertainties follow a Gaussian distribution) at $t=6$: (a) $\epsilon=0.05$; (b) $\epsilon=0.10$; (c) $\epsilon=0.15$; (d) $\epsilon=0.25$.}
	\label{fig_uncertaintySet_case3}
\end{figure}

Fig. \ref{fig_uncertaintySet_case3} presents the results of uncertainty sets constructed by different models with normally distributed uncertainties. Similar to \textbf{Cases 1} and \textbf{2}, the proposed uncertainty set has the smallest volume. The other two are relatively large because they must cover some extreme samples (such as at the bottom-right corner).

\subsubsection{Optimality, feasibility and time-efficiency}

\begin{figure}
	\centering
	 			\vspace{-4mm}
	\includegraphics[width=1\columnwidth]{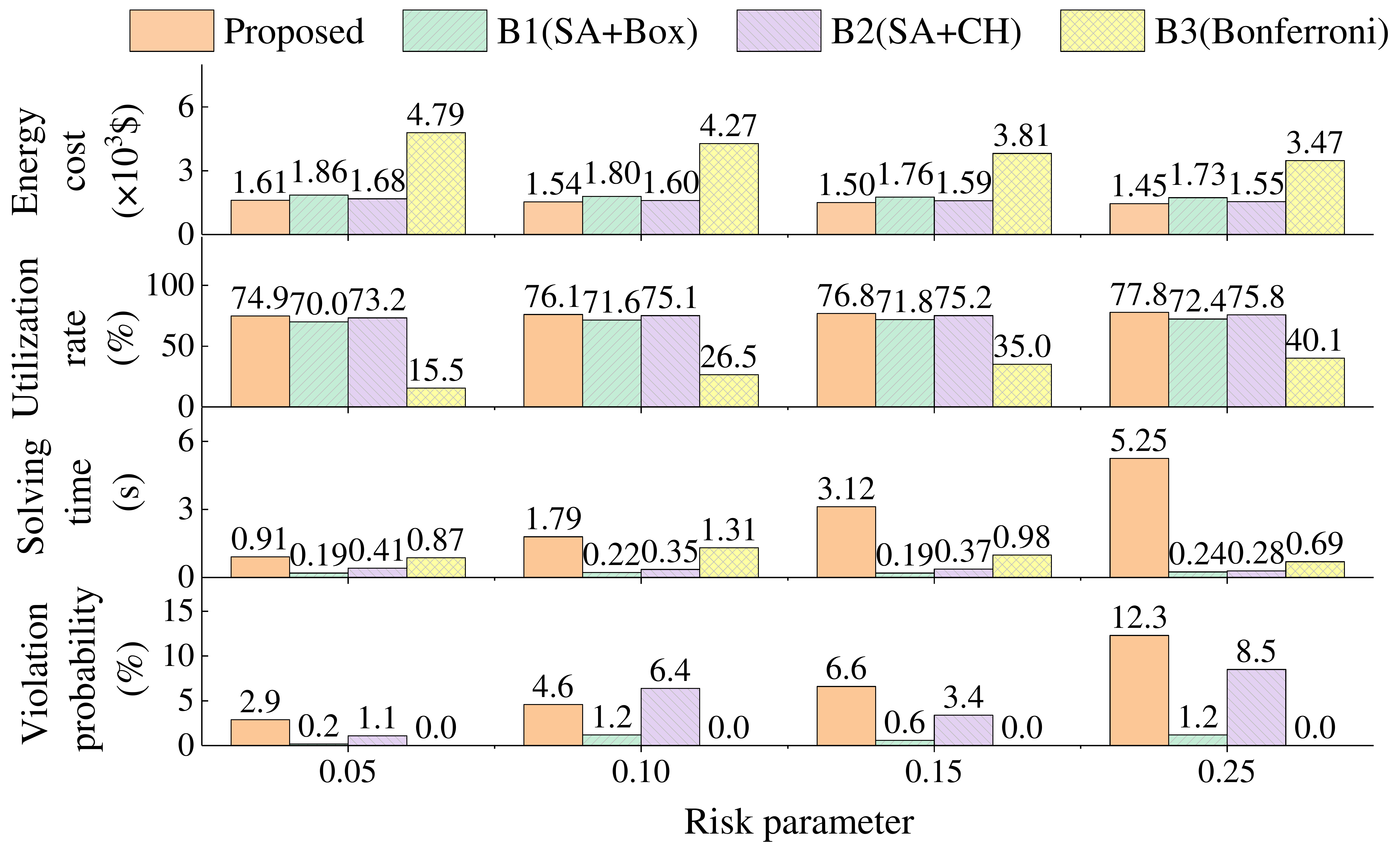}\vspace{-4mm}
	\caption{Results of energy costs, average utilization rates of DRG, solution times, and maximum violation probabilities obtained by different models in \textbf{Case 3} (uncertainties follow a Gaussian distribution).
	}
	\label{fig_Gaussian}
	 		\vspace{-4mm}
\end{figure}
Fig. \ref{fig_Gaussian} summarizes the energy cost, average utilization rate of DRG, solution time, and maximum violation probability. The energy cost of the proposed method is the lowest, e.g.,  13.4\%, 4.2\%, and 66.4\% less than those of \textbf{B1}--\textbf{B3}, respectively, with risk parameter $\epsilon=0.05$. Its utilization rate of DRG is also the highest, e.g., 4.9\%, 1.7\%, and 59.4\% higher than those of \textbf{B1}--\textbf{B3}, respectively, at $\epsilon=0.05$. Although its solution time is greater, the computational efficiency is still acceptable.

The results in \textbf{Cases 1}--\textbf{3} also confirm that the proposed method can handle JCCs with arbitrarily distributed uncertainties, confirming its generalization performance.

\begin{figure}
	\centering
	\vspace{-4mm}
	\includegraphics[width=0.95\columnwidth]{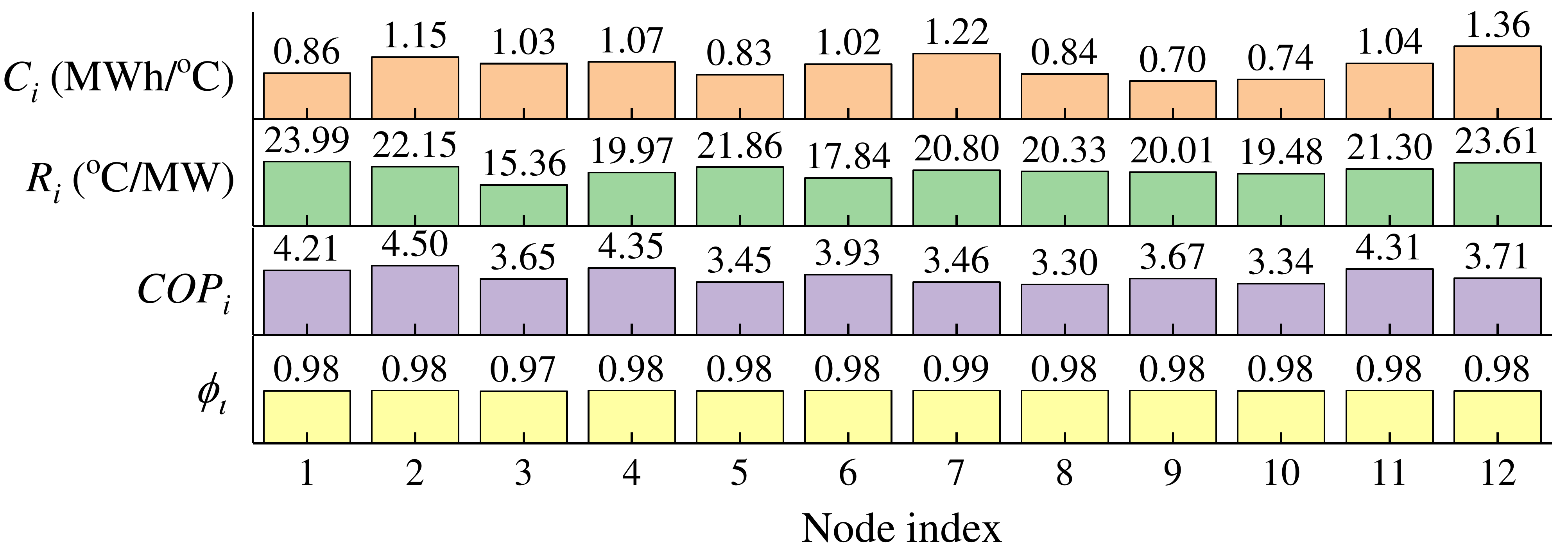}
		\vspace{-4mm}
	\caption{{The heterogeneous parameters in different nodes.}}
	\label{fig_parameter_heter}
	 		\vspace{-4mm}
\end{figure}

{
\subsection{Effectiveness under heterogeneous parameters}
In \textbf{Cases 1--3}, we assume that the building parameters of all nodes are homogeneous. To better verify the performance of the proposed model, we implement a  case study in which different nodes have heterogeneous parameters. Fig. \ref{fig_parameter_heter} shows the heat capacity  $C_i$ of the building, thermal resistance $R_i$ between indoor and outdoor environments, coefficient of performance $\text{COP}_i$, and coefficient $\sqrt{(1 - \phi_i^2)}/\phi_i$ used in (\ref{eqn_q_HV}). Buildings connected to different nodes obviously have heterogeneous parameters and various demands.} 

\begin{figure}
	\centering
	\includegraphics[width=1 \columnwidth]{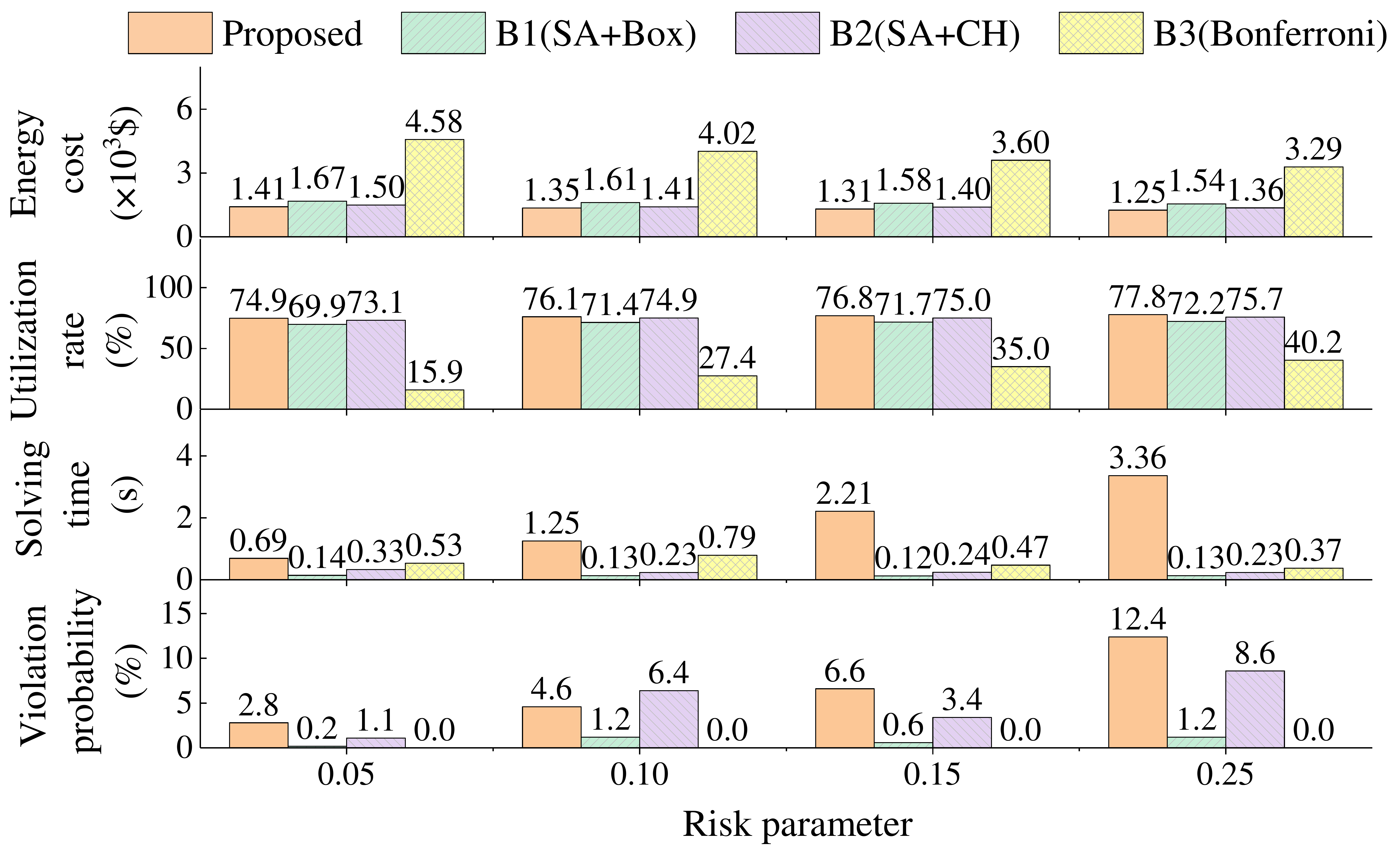}\vspace{-4mm}
	\caption{{Simulation results in the case with heterogeneous parameters.}}
	\label{fig_result_heter}
	 		\vspace{-2mm}
\end{figure}

{
Fig. \ref{fig_result_heter} summarizes the results of this case study. Even with some heterogeneity in the parameters, the proposed method achieves the highest energy efficiency and utilization rate of DRG. Moreover, the violation probability of the JCC is always less than the given risk parameter. Although the solution time is longer, the maximum solution time is only 3.36 s.}

\subsection{Effectiveness of flexibility from HVAC loads}
\begin{figure}
		\vspace{-4mm}
	\centering
	\subfigure[]{\includegraphics[width=0.49\columnwidth]{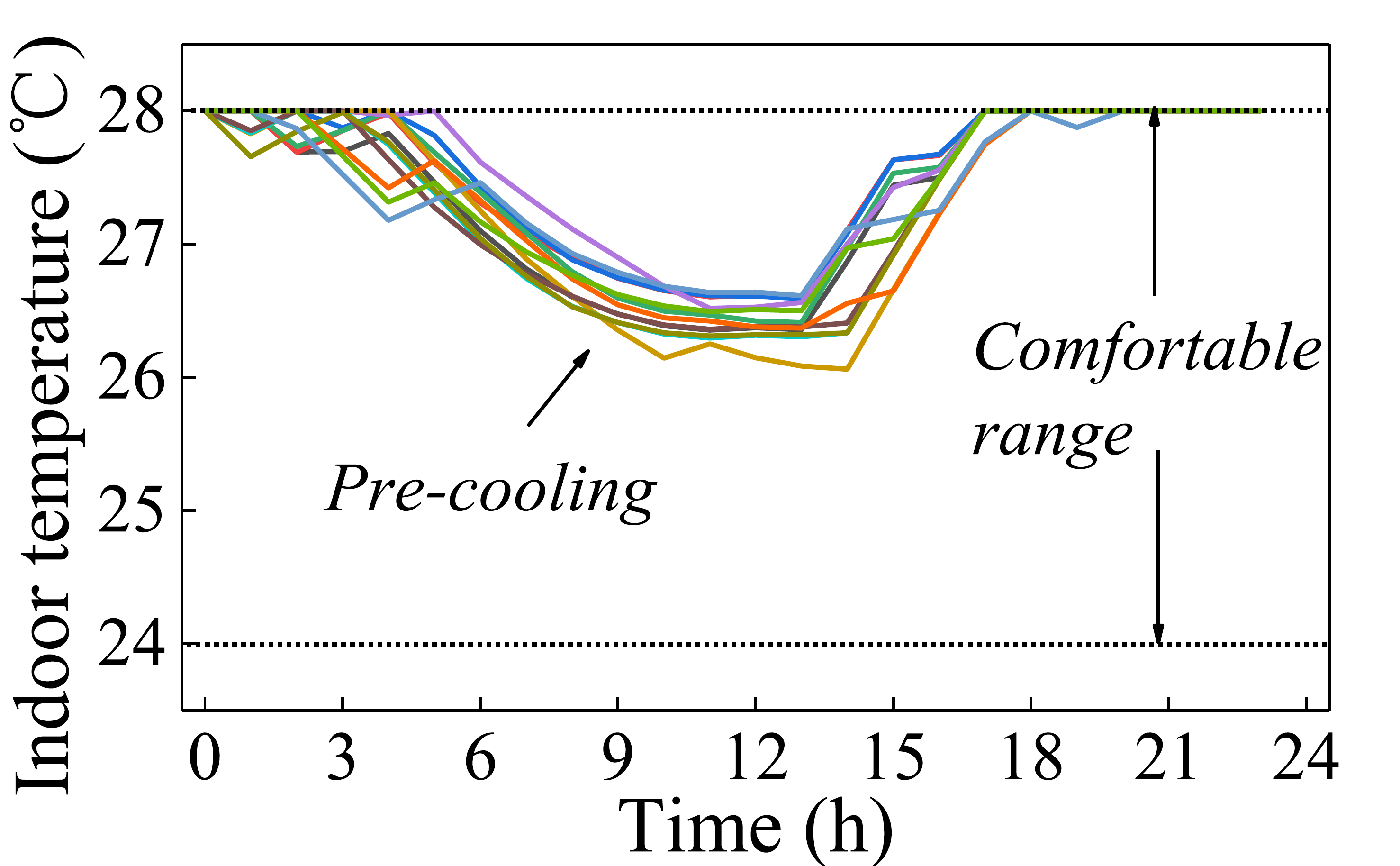}}
	\subfigure[]{\includegraphics[width=0.49\columnwidth]{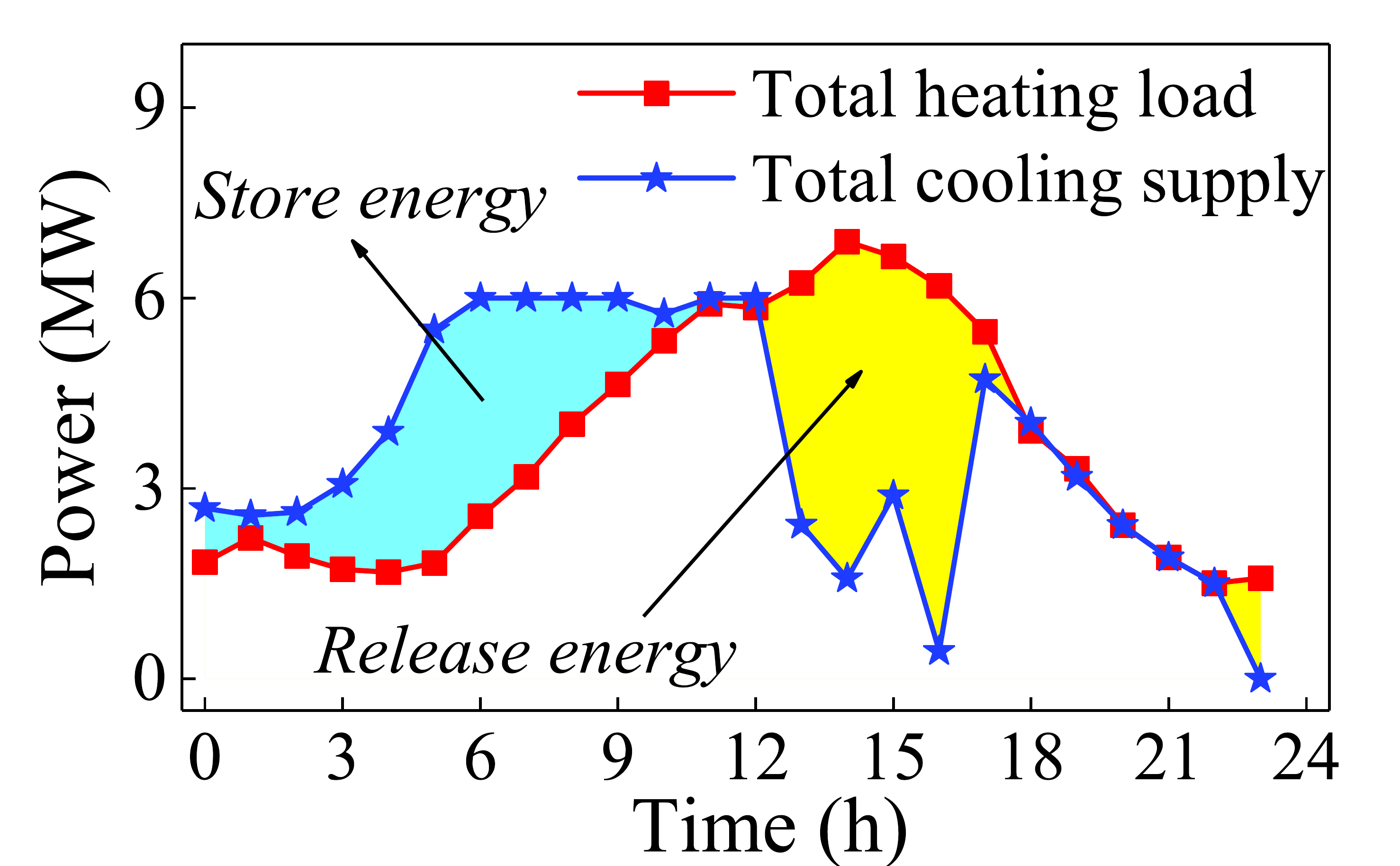}}
	\vspace{-4mm}
 	\caption{Results of (a) indoor temperatures and (b) total heating load and cooling supply with Beta distributed uncertainties and $\epsilon=0.05$. }
	\label{fig_thermal}
	\vspace{-4mm}
\end{figure}

Fig. \ref{fig_thermal} shows the indoor temperatures, total heating load (heat loads from indoor sources plus heat transfer from outdoor environments), and cooling supply. The results are obtained by the proposed model with Beta distributed uncertainties with $\epsilon=0.05$. During the period from 4:00 a.m. to 12:00 a.m., a significant temperature drop occurs in each building, which we refer to as pre-cooling, and the energy storage is based on this. During these hours, the available DRG power is relatively high, and the electricity price is low. To reduce the whole-day energy cost, the distribution network unlocks building thermal flexibility to store as much cooling power as possible for later use. Thus, the total cooling supply is much higher than the required total heating load during this period, as shown in Fig. \ref{fig_thermal}(b). In the next few hours, the electricity price increases while the available DRG power decreases, and the stored cooling power is released. Therefore, the total cooling supply is visibly lower than the total heating load, leading to a noticeable increase in indoor temperatures. These results confirm that buildings can serve as batteries to provide operational flexibility for distribution networks.

\begin{figure}
		\vspace{-4mm}
	\centering
    \includegraphics[width=0.8\columnwidth]{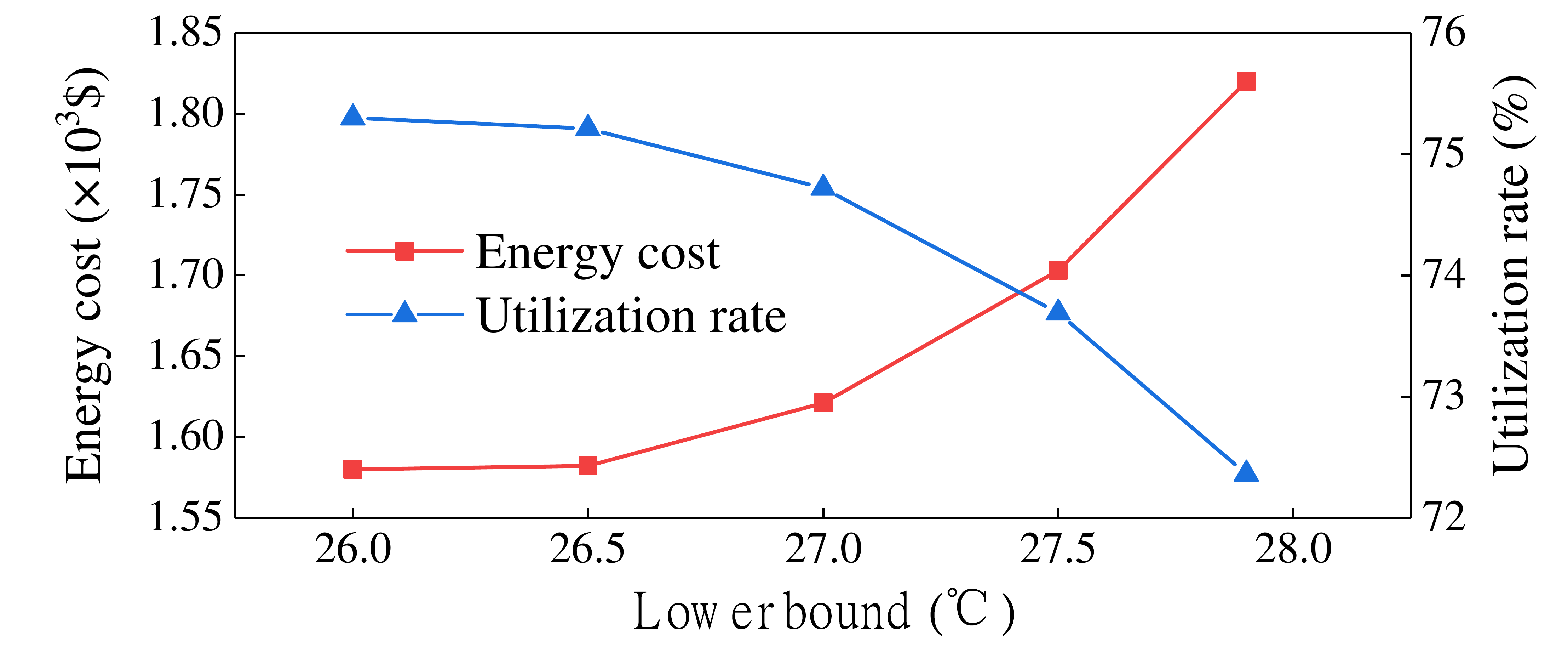}
    \vspace{-4mm}
 	\caption{Effects of temperature lower bound on energy cost and average utilization rate of DRG with $\epsilon=0.05$.}
	\label{fig_flexibility}
	\vspace{-4mm}
\end{figure}

Since the energy storage of buildings is based on pre-cooling, the variation of the lower bound of the indoor temperature, i.e., $\underline{\bm \theta}$, influences the capacity of the building's thermal flexibility. We change $\underline{\bm \theta}$ to   verify the effectiveness of this flexibility, with results as shown in Fig. \ref{fig_flexibility}. With the growth of $\underline{\bm \theta}$, the capacity of the building thermal flexibility becomes smaller because less cooling power can be stored in indoor environments, leading to a lower utilization rate of DRG and higher energy cost.

\section{Conclusions} \label{sec_conclusion}
This paper proposes a learning-based approach to promote DRG integration by scheduling flexible HVAC loads. The violation probabilities of all critical constraints are managed by JCCs to ensure the security of the whole system. To overcome the intractability of JCCs, this paper develops a robust constraint with a novel OC-SVC-based polyhedron uncertainty set to safely approximate JCCs. A linear counterpart was developed to guarantee computational efficiency. Numerical experiments on an IEEE 13-bus system indicated that the proposed uncertainty set can tightly cover most historical samples. Hence, the proposed method could achieve better optimality than the widely used scenario approach and Bonferroni approximation. 
Simulation results also confirm that the flexibility of HVAC loads can promote the integration of DRG and enhance energy efficiency.

\appendices
\setcounter{table}{0}   
\renewcommand\thetable{\Alph{section}\arabic{table}} 
\section{} \label{app_1}
\emph{Proof of \textbf{Proposition} \ref{prop_2}}:
According to Table \ref{tab_A1}, the radius $R$ is equal to the distance from the vector $\bm o$ to any boundary support vector,
\begin{align}
    R^2 = &\Vert \bm \phi(\bm \xi^{(k)}) - \bm o \Vert^2, \quad k \in \text{BSV}. \label{eqn_R_1}
\end{align}
Based on (\ref{eqn_KKT1}), we can substitute $\bm o=\sum_{n \in \mathcal{N}}\alpha_n \bm \phi(\bm \xi^{(n)})$ in  (\ref{eqn_R_1}) to obtain 
\begin{align}
\begin{split}
R^2 =&K(\bm \xi^{(k)},\bm \xi^{(k)})-2\sum_{n \in \mathcal{N}}\alpha_nK(\bm \xi^{(k)},\bm \xi^{(n)})\\
&+\sum_{n \in \mathcal{N}}\sum_{m \in \mathcal{N}}\alpha_n\alpha_mK(\bm \xi^{(n)},\bm \xi^{(m)}), k \in \text{BSV}. 
\end{split}\label{eqn_R}
\end{align}
Similarly, by substituting $\bm o=\sum_{n \in \mathcal{N}}\alpha_n \bm \phi(\bm \xi^{(n)})$, the region bounded by the minima sphere, i.e., $\left\lbrace \bm \xi \left|\ \Vert \bm \phi(\bm \xi) - \bm o \Vert^2 \leq R^2\right. \right.\rbrace$, can be expressed as
\begin{align}
\begin{split}
\mathcal{U} = & \left\lbrace \bm \xi \left|K(\bm \xi, \bm \xi) - 2\sum_{n \in \mathcal{N}}\alpha_nK(\bm \xi,\bm \xi^{(n)}) \right. \right. \\
& \left. +\sum_{n \in \mathcal{N}}\sum_{m \in \mathcal{N}}\alpha_n\alpha_mK(\bm \xi^{(n)},\bm \xi^{(m)}) \leq R^2 \right \rbrace . \label{eqn_uncertainty_set}
\end{split} 
\end{align}
Then,  substituting  (\ref{eqn_R}) and (\ref{eqn_kernel}) in  (\ref{eqn_uncertainty_set}), we can rewrite the uncertainty set   as
\begin{align}
\mathcal{U} =& \left\lbrace \bm \xi \left|\sum_{n \in \mathcal{N}}\alpha_nK(\bm \xi,\bm \xi^{(n)})  \geq \sum_{n \in \mathcal{N}}\alpha_nK(\bm \xi^{(k)},\bm \xi^{(n)}) \right. \right\rbrace \notag \\
=& \left\lbrace \bm \xi \left|\sum_{n \in \text{SV}}\alpha_n\Vert\bm W (\bm \xi-\bm \xi^{(n)}) \Vert_1  \leq \gamma \right.  \right\rbrace, \label{eqn_uncertainty_set_11}
\end{align}
where   $\gamma=\sum_{n \in \text{SV}}\alpha_n\Vert \bm W(\bm \xi^{(k)} - \bm \xi^{(n)})\Vert_1$. Introducing   auxiliary variable $\bm \upsilon_n$, the L1-norm operator can be eliminated, and $\mathcal{U}$ can be reformulated as (\ref{eqn_uncertainty_set2}) in \textbf{Proposition} \ref{prop_1}.

\section{} \label{app_2}
\emph{Proof of \textbf{Proposition} \ref{prop_2}}: The LHS term in  (\ref{eqn_RO_2}) is equivalent to the following linear programming problem:
\begin{align}
&\max_{\bm \xi_t, \bm \upsilon_{n,t}} \quad (\bm H_{m,t} \bm \xi_t)^\intercal \bm y_{t}, \tag{$\textbf{P-A1}$}\\
&\begin{array}{r@{\quad}r@{}l@{\quad}l}
\text{s.t.} && \sum_{n \in \text{SV}_t}\alpha_{n,t} \bm \upsilon_{n,t}^\intercal \bm 1 \leq {\gamma}_{t}, \\
&&-\bm \upsilon_{n,t} \leq \bm W(\bm \xi_t-\bm \xi_{t}^{(n)})\leq \bm \upsilon_{n,t}, \quad \forall n \in \text{SV}_t.
\end{array} \notag
\end{align}
By introducing   Lagrange multipliers $\bm \rho_{n,t}$, $\bm \mu_{n,t}$, and $\pi_{t}$, we can obtain its dual problem: 
\begin{align}
&\min_{\substack{\bm \rho_{n,m,t},\\ \bm \mu_{n,m,t},\\ \pi_{m, t}}} \quad \sum_{n \in \text{SV}_t}(\bm \mu_{n,m,t} - \bm \rho_{n,m,t})^\intercal \bm W_t \bm \xi_t^{(n)} + \pi_{m,t} {\gamma}_{t},  \tag{$\textbf{D-A1}$}\\
&\begin{array}{r@{\quad}r@{}l@{\quad}l}
\text{s.t.} && \sum_{n \in \text{SV}_t}\bm W_t (\bm \rho_{n,m,t} - \bm \mu_{n,m,t}) + \bm H_{m,t}^\intercal \bm y_{t} = \bm 0, \\
&&\bm \rho_{n,m,t} + \bm \mu_{n,m,t} = \pi_{m,t} \alpha_{n,t} \bm 1,\\
&&\pi_{m,t}\geq 0, \quad \bm \mu_{n,m,t},\bm \rho_{n,m,t} \in \mathbb{R}^D_+,\quad \forall n \in \text{SV}_t. 
\end{array} \notag
\end{align}
Note that \textbf{P-A1} is a linear problem, so the optimal solutions of \textbf{P-A1} and \textbf{D-A1} are equal based on strong duality.
Finally, by substituting \textbf{D-A1} in  (\ref{eqn_RO_2}), \textbf{Proposition} \ref{prop_2} can be proved.
 
\footnotesize
\bibliographystyle{ieeetr}
\bibliography{ref}
\end{document}